\documentclass[a4paper,11pt]{article}
\pdfoutput=1 % if your are submitting a pdflatex (i.e. if you have
\usepackage{jinstpub} % for details on the use of the package, please
\usepackage{siunitx}
\usepackage{subfig}
\usepackage{appendix}
\title{\boldmath Secondary Electron Emission from Multi-layered TiN/Al\textsubscript{2}O\textsubscript{3} Transmission Dynodes}

\author[a,b,1]{H.W. Chan,\note{Corresponding author.}}
\author[a,b]{V. Prodanovi\'c,}
\author[c]{A.M.M.G. Theulings,}
\author[c]{C.W. Hagen,}
\author[b]{P.M. Sarro}
\author[a]{and H. v.d Graaf}

\affiliation[a]{National Institute for Subatomic Physics (NIKHEF),\\Science Park 105, 1098 XG, Amsterdam, The Netherlands}
\affiliation[b]{Faculty of Electrical Engineering, Mathematics, and Computer science, Department of microelectronics/ECTM,\\Feldmannweg 17, 2628 CT, Delft, The Netherlands}
\affiliation[c]{Faculty of applied sciences, Department of Imaging Physics, Delft University of Technology,\\Lorentzweg 1. 2628 CJ, Delft, The Netherlands}

\emailAdd{h.w.chan@hotmail.com}

\abstract{The (secondary) electron emission from multilayered Al\textsubscript{2}O\textsubscript{3}/TiN membranes has been investigated with a hemispherical collector system in a scanning electron microscope for electrons with energies between \SI{0.3} and \SI{10}{\keV}. These ultra-thin membranes are designed to function as transmission dynodes in novel vacuum electron multipliers. Two different types, a bi-layer and a tri-layer, have been manufactured by means of atomic-layer deposition (ALD) of aluminum oxide and sputtering of titanium nitride. The reflection and transmission electron yield ($\sigma_R$, $\sigma_T$) have been measured for both types of membranes. In comparison, the tri-layer membranes outperformed the bi-layer membranes in terms of transmission electron yield for films with the same effective thickness. The highest transmission electron yield was measured on an Al\textsubscript{2}O\textsubscript{3}/TiN/Al\textsubscript{2}O\textsubscript{3} film with layer thicknesses of 5/2.5/5 \SI{}{\nm}, which had a maximum transmission electron yield $\sigma_\text{T}^\text{max}(E_0)$ of $3.1(\SI{1.55}{\keV})$. Furthermore, the bi-layer membranes have been investigated more in-depth by performing an additional measurement using a positive sample bias to separate the transmitted fraction $\eta_T$ and the transmission secondary electron yield $\delta_T$. The transmitted fraction was used to determine the transmission parameter $p$, which characterizes the interaction of primary electrons (PEs) in thin films. The transmission secondary electron yield was used to compare the energy transfer of PEs in films with different thicknesses.}

\keywords{secondary electron emission; transmission dynode; photomultiplier; vacuum electron multipliers; atomic layer deposited aluminum oxide; titanium nitride; ultra-thin films}

\arxivnumber{2008.08997} % only if you have one

\begin{document}
\maketitle
\flushbottom

%%%%%%%%%%%%%%%%%% Introduction %%%%%%%%%%%%%%%%%%

\section{Introduction}
\label{sec:intro}
\subsection{Novel vacuum electron multipliers}
\label{ssec:novel}

%%% vacuum electron multipliers %%%

\begin{figure}
\centering 
\includegraphics[width=0.8\textwidth,origin=c,angle=0]{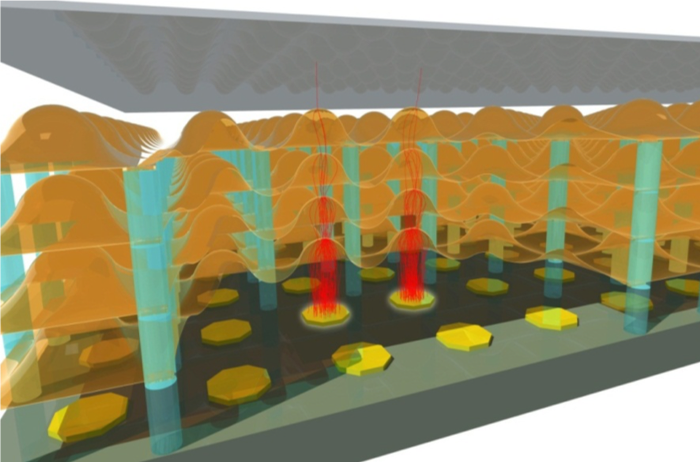}
\caption{\label{fig:1} The Timed Photon Counter consists of a traditional photocathode, a stack of tynodes and a TimePix chip within a compact vacuum enclosure. The electric potential between each tynode in the stack increases step-wisely from top to bottom. A soft photon can emit a photoelectron from the photocathode, which is then accelerated towards the first tynode due to the electric field. On impact, the incoming electron generates multiple transmission secondary electrons (TSEs) from the ultra-thin tynode, which escape from the backside. The concave surface of the tynode bundles and accelerates the TSEs towards the next tynode. The multiplication process repeats $N$ times for each layer. Eventually a number of $\sigma_T^N$ electrons appear above the individual pixel and are detected by the digital circuitry of the TimePix chip.}
\end{figure}

Vacuum electron multipliers, such as photomultiplier tubes (PMTs), employ secondary electron emission (SEE) for photon detection \cite{Hamamatsu2007}. The detection principle is the conversion of photons into photoelectrons by the photoelectric effect and subsequent electron multiplication in vacuum. A photoelectron, accelerated from the photocathode towards the first dynode, will generate multiple secondary electrons (SEs) upon impact. The SEs are then guided and accelerated towards the next dynode.  As they traverse from dynode to dynode, their number increases, until the SEs are collected by the anode.

PMTs are one of the most sensitive photon detectors and are still widely used for single-photon detection due to its high gain, low noise and large acceptance surface. Though, there are a few disadvantages to the design. First, the time resolution in the order of a nanosecond is poor in comparison with silicon photomultipliers with single-photon avalanche diodes \cite{Donati2014}. The time resolution depends on the spread in transit times of the SEs in the dynode section of a PMT. Also, the SEs are susceptible to magnetic fields, which exclude PMTs to be used in applications with strong magnetic fields. And lastly, the dynode stack makes PMTs voluminous, fragile and expensive.
	
The goal of the MEMBrane project is to develop a vacuum electron multiplier that outperforms PMTs in terms of time and spatial resolution \cite{VanderGraaf2017}. The device, the Timed Photon Counter (TiPC), has the same detection principle as a PMT, but has transmission dynodes (tynodes) as multiplication stages instead of (reflective) dynodes (Figure \ref{fig:1}). Tynodes are extremely thin membranes where an impinging primary electron (PE) on the frontside releases multiple secondary electrons from the backside. This distinctive property allows tynodes to be closely stacked on top of each other. The distance between neighbouring stages, in comparison with dynodes in PMTs, is greatly reduced and the electric field is nearly homogenous. As a result, the time resolution improves: the pulse width and the rise time of the signal will be smaller due to the more uniform and shorter electron paths. In addition, the susceptibility to magnetic fields is reduced due to the increased electric field strength. In terms of spatial resolution, 2D spatial information is gained by combining the planar tynode stack with a CMOS-pixelchip (TimePix) as read-out. Lastly, TiPC is a smaller and more compact photodetector in comparison with a PMT.

A tynode is an ultra-thin membrane that (1) consist of a material with a high secondary electron yield (SEY), (2) is mechanically strong and (3) is electrically conductive. The transmission secondary electron yield (TSEY) is defined as the ratio between the number of incoming PE, with an energy $E_0$, and the number of outgoing SEs in transmission. For TiPC, the goal is to achieve a TSEY of 4 or higher for primary electrons with sub-2 \SI{}{\keV} energy. As such, mechanically strong and thin membranes are required, since the range of PEs is energy dependent. \emph{In-plane conductivity} is required to replenish the emitted electrons in the irradiated region of the membrane by providing an electrical path to the conductive mesh in which the membranes are suspended, otherwise charge-up effects will occur during prolonged electron irradiation. 

%%% TSEE %%%

\subsection{Transmission secondary electron emission}
\label{ssec:theory}

\begin{figure}
\centering 
\includegraphics[width=0.8\textwidth,origin=c,angle=0]{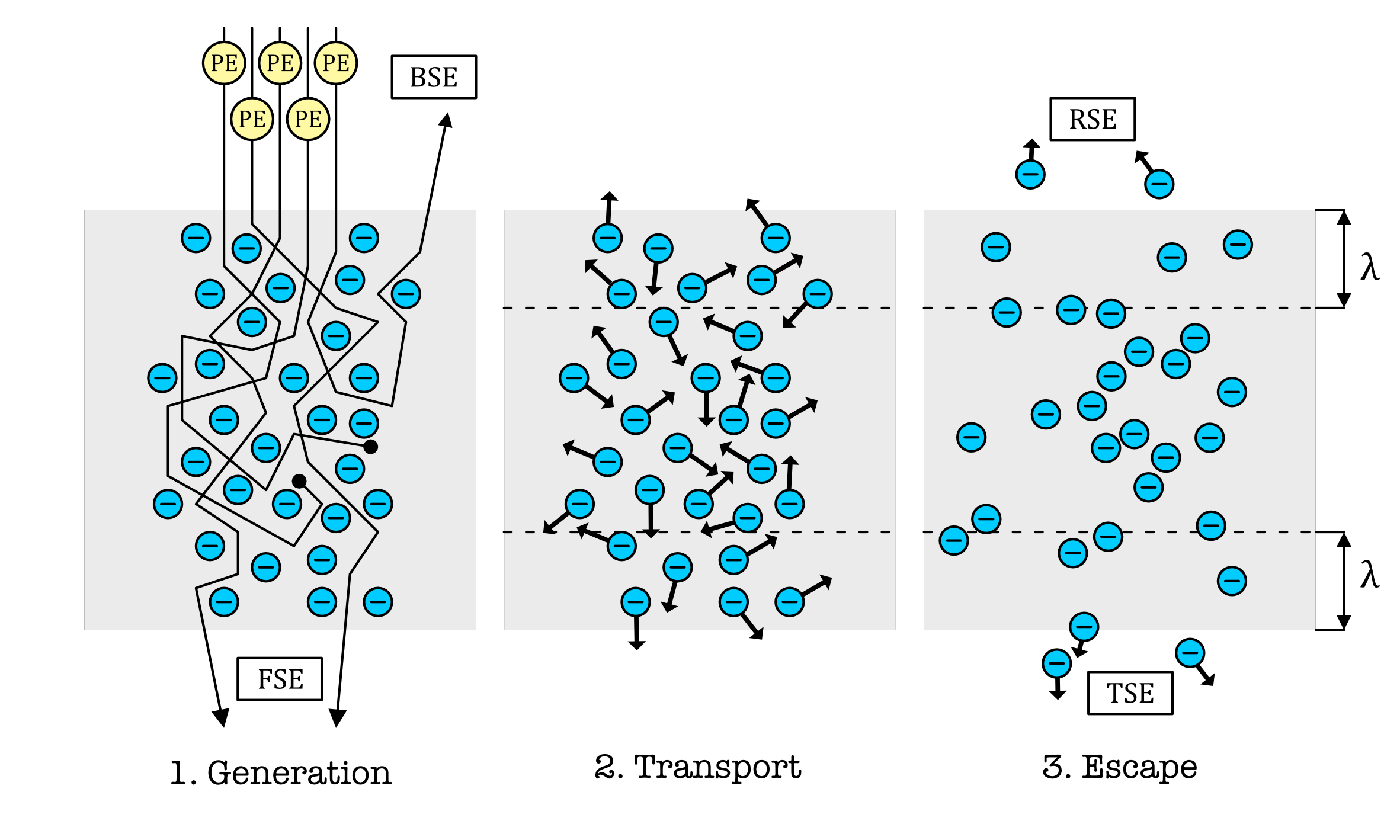}
\caption{\label{fig:2} Three-step model of SE generation. The three steps are treated independently in the elementary theory of SEE. The first step describe energy transfer of PEs in the film/bulk. The second step models the transport of internal SEs. The third step describe the escape probability of SEs from the material into vacuum.}
\end{figure}

Secondary electron emission is described as a three-step process: \emph{generation}, \emph{transport} and \emph{escape} of internal SEs \cite{Bruining1954,Dekker1958}. This model can be extended to thin membranes by including the exit surface of the membrane in transmission (Figure \ref{fig:2}). The first step of the model treats PE interaction, energy transfer and SE generation. A PE that interacts within a thin membrane will scatter and lose energy. Some of the energy is used to generate internal SEs. The PE itself can be reflected, absorbed or transmitted by the membrane. Reflected PE are designated as backscattered electrons (BSE), while transmitted electrons as forward-scattered electrons (FSE). They are distinguished from SEs by their energy, which is $E >$ \SI{50}{\eV}. The second step describes the transport of internal SEs within the material. The band gap model is used to explain the difference in transport in metal, semiconductors and dielectrics \cite{Bruining1954,Dekker1958}. The wide band gap of dielectrics allows SEs that are promoted to the conduction band to travel a relatively large distance with few interactions. This increases the probability of the SEs to reach the surface. The third step models the escape of internal SEs into vacuum at the solid-vacuum boundary. Internal SEs with sufficient energy to overcome the work function or electron affinity can escape into vacuum. Only internal SEs that are generated near the surface have a chance to escape. The escape probability is given as an exponential decay function with $\lambda$ the mean free path of SEs. The secondary electrons that escape from the entrance are designated as reflection secondary electrons (RSE) and from the exit as transmission secondary electrons (TSE). 

The reflection secondary electron yield (RSEY) of a surface depends on its material properties and surface condition. In general, dielectrics have higher yields in comparison with semiconductors and metals \cite{Bruining1954,Dekker1958}. This is attributed to the wide band gap of dielectrics which benefits the transport of internal SEs. Surface treatment, such as caesiation and hydrogen-termination, can lower the electron affinity, which will also increase the escape probability of internal SEs. In some cases, even negative electron affinity (NEA) can be achieved; an internal SE that reaches the surface will encounter no barrier and will be pushed into vacuum. This is beneficial for SEE. The total reflection electron yield (REY) of C(100) diamond, for instance, increased from 3 to 60 and 132 by Cs- and H-termination respectively \cite{Yater2000}. 

For transmission SEE, the thickness of the membrane is an additional parameter that affects the total transmission electron yield (TEY). The onset of transmission SEE is expected to occur when PEs are expected to penetrate through the tynode. This characteristic is defined as the critical energy $E_c$ for which 1\% of the PEs manages to pass through: $\eta_T(E_c) = 0.01$ \cite{Fitting1974}. The coefficient $\eta_T$ is the FSE coefficient or the transmitted fraction. A second characteristic (tied to the thickness) is the energy $E_{T}^{\text{max}}$ at which the maximum TEY $\sigma_{T}^\text{max}$ is achieved: $\sigma_{T}^\text{max}(E_{T}^{\text{max}})$. Both are unique defining features of a TEY curve correlated to the thickness of the membrane. 

The distance that a PE with energy $E_0$ can travel is defined as the range and is given by $R = C {E_0}^n$, where $C$ is a constant that is material dependent and $n$ a constant that depends on the energy-range of the PEs \cite{Kanaya1972}. There are a variety of range-energy relations \cite{Lukiyanov2009}. The accuracy of these relations depends on the material considered and the energy of the PE. For sub-10 \SI{}{\keV} electrons and alumina as material, the range-energy relation given by Fitting \cite{Fitting1974} is the most accurate and is given by
\begin{equation}
\label{eq:5}
R = 90\rho^{-0.8}E_0^{1.3}
\end{equation}
where $R$ is the range in \SI{}{\nano\meter}, $\rho$ the density given in \SI{}{\gram\per\cm^3} and $E_0$ the primary beam energy in \SI{}{\keV}. The range $R$ of a PE in different materials will differ, which makes comparison of composite films to single-material films difficult. However, an effective layer method can be applied to films with different materials \cite{Fitting2007}. The contribution to the stopping power of material 2 can be replaced by material 1 with an effective layer thickness given by	
\begin{equation}
\label{eq:6}
d_1^{\text{eff}} = \biggl(\frac{d_1}{R_1}\biggr) ^{p_1/p_2}R_2
\end{equation}
where $p_{1,2}$ is the transmission parameter and $R_{1,2}$ is the range in the first and second material respectively. The total effective film thickness is then given by $d = d_1^{\text{eff}} + d_1$.
\\

%% SEY measurement %%

A basic SEE measurement setup consists of an electron gun, a sample holder and an electron collector, which are placed inside a vacuum chamber. The standard procedure is to measure the SE current from the surface and compare it to the PE current. Depending on the method, the ratio is either the total reflection electron yield (REY) or reflection secondary electron yield (RSEY). In literature, the distinction between RSEs and BSEs is often not made and one can assume that their combined yield is reported. Also, the reported REY of the same material varies widely, which is attributed to the different experimental conditions \cite{Bruining1954,Dekker1958}. For instance, the surface condition plays an important role as was shown on alkali halide single crystals by Whetten \cite{Whetten1964}. The crystals were cleaved in vacuum and the pristine surface had a higher REY. SEE measurements are often performed in dedicated ultra-high vacuum chambers with the option of surface and/or thermal treatments. 

SEE measurements on dielectrics are more challenging due to the build-up of charge on the surface \cite{Chvyreva2014}. The recommended strategy is to limit the electron dose, which can be achieved by using a pulsed electron gun \cite{Chvyreva2014,Whetten1964} and/or to neutralize the charge with a flood gun between measurements \cite{Handel1966}. A different approach is to determine the REY by measuring the surface charge using the Kelvin probe method \cite{Balcon2012}.

For transmission SEE measurement, a dual collector setup can be used to measure the REY and TEY \cite{Fitting1974}. The disadvantage of this 'closed' system is the difficulty of targeting the sample surface. If the membrane size is extremely small, a part of the electron beam will irradiate the inactive surface. Therefore, a modified collector method was developed to determine the transmission yield within a scanning electron microscope (SEM) \cite{Prodanovic2018}. The imaging capability of the SEM was used to locate and to direct the electron beam on the ultra-thin membranes. Despite using a continuous beam, charge-up effects were not observed on films/membranes on which a conductive TiN layer was sputtered.

A caveat of this method is the lack of ultra-high vacuum in the SEM, which operates at \SI{1e-6}{\milli\bar} instead of \SI{1e-9} {\milli\bar} or lower. As a result, surface contamination can form after prolonged surface irradiation \cite{Vladar2001}. The contamination rate depends on the electron dose per unit surface and can be lowered by scanning the electron beam over the surface. A comparison between this setup and a dedicated ultra-high vacuum system have been made by measuring the reflection SEY of a SiN and an Al\textsubscript{2}O\textsubscript{3}  film \cite{VanderGraaf2017}. The results were in good agreement and contamination effects were not observed. However, dedicated surface termination studies should be performed in ultra-high vacuum systems.

%% Transmission dynode %%

\subsection{Transmission dynode}
\label{ssec:state}

In the past, different groups of materials were considered as tynode material, such as alkali halides, semiconductors and diamond. One of the first working transmission-type photomultipliers has been built by Sternglass \& Wachtel. \cite{Sternglass1955,Sternglass1956}. The tynodes consist of porous potassium chloride (KCl) deposited on top of an aluminum foil. The high TEY of porous materials is due to the build-up of charge inside the pores of the material, which results in a strong electric field where (secondary) electrons are accelerated internally causing an avalanche type of SE emission. The typical inter-stage operating voltage is \SI{5}{\keV} with a maximum TEY $\sigma_{T}^\text{max}$ of 8. Despite the high TEY, the required high voltage for a multi-stage device limits its applicability. Also, the lifetime of the devices was poor and further research in the aging mechanism was needed \cite{Sapp1964}. The TEY of other alkali halides (CsI, KCl, NaF and LiF) have been measured by Llacer \& Garwin \cite{Llacer1969}. They were deposited onto an Al/Al\textsubscript{2}O\textsubscript{3} membrane as support, which increased the overall film thickness. The highest TEY of 8 (\SI{8}{\keV}) was measured for cesium iodide. The best performing alkali halide was reported by Hagino et al. on caesium activated CsI. They achieved a TEY of 27 (\SI{9}{\keV}) for Al\textsubscript{2}O\textsubscript{3}/Al/CsI(Cs) films \cite{Hagino1972}. A second group of materials that was considered were semiconductors, such as silicon and gallium arsenide, that benefitted from negative electron affinity (NEA). A TEY of 725 (\SI{25}{\keV}) for a 4-5 \SI{}{\micro\meter} thick silicon film with NEA was achieved by Martinelli \cite{Martinelli1970}. More recently, various types of diamond have been studied as SEE materials for transmission dynodes \cite{Yater2002,Yater2003,Yater2004,Yater2011}. The highest TEY of 5 (\SI{7}{\keV}) was obtained for nano-crystalline chemical vapor deposition diamond. Although the results are promising, it is unclear whether thinner nano-crystalline diamond can be manufactured with the same quality, since the growth process requires nucleation sites. 

The large PE energy that is required to achieve the high TEY makes these tynodes impractical. Ideally for TiPC, the tynodes need to perform optimally for sub-2 \SI{}{\keV} electrons. The limiting factor is thus the film thickness. After a thorough review, it became clear that the tynode needs to be self-supporting \cite{Tao2016}. The choice of materials is therefore limited to materials that are mechanically strong and have a high SEY. Accordingly, we approached the problem from a micro-fabrication/engineering point of view. Silicon nitride tynodes were fabricated by low-pressure chemical vapor deposition and aluminum oxide tynodes by means of atomic layer deposition (ALD) \cite{Prodanovic2015,Prodanovic2018}. Monte-Carlo simulation has shown that the optimum thickness for aluminum oxide tynodes is about \SI{10}{\nm} \cite{VanderGraaf2017}. Therefore, the ultra-thin membranes, with a diameter of 10 – 30 \SI{}{\micro\meter}, were suspended within a supporting mesh with an array of 64-by-64 small windows \cite{Prodanovic2018}. A TEY of 1.57 (\SI{2.85}{\keV}) was measured for TiN/SiN films and a TEY of 2.6 (\SI{1.45}{\keV}) for TiN/Al\textsubscript{2}O\textsubscript{3} films. 

Titanium nitride was chosen as a conductive layer to provide in-plane conductivity. The added layer does increase the thickness, but has a relatively low stopping power due to the low $Z$ value of TiN. Other conductive materials were considered, such as metals (Al, Cr), but they will most likely oxidize during the fabrication process, whereas TiN is chemically inert in ambient conditions \cite{Logothetidis1999}. Charge-up within the alumina layer was not observed, i.e. the emission current is constant during exposure. The mechanism that provides \emph{normal-to-the-plane} conductivity from the conductive layer to the charged region in the dielectric film can be either explained by electron-beam induced current (EBIC) \cite{Taylor1979} and/or electron tunnelling \cite{Lee2001}. \\

In this paper we will determine and compare the (transmission) SEE of two types of multilayered membranes: a bi- and a tri-layer. The conductive layer of the bi-layer TiN/Al\textsubscript{2}O\textsubscript{3} membrane is deposited after releasing the membrane. Due to the topography of the surface, the sputtered TiN layer is less uniform, which increases the risk of a disconnected layer. The fabrication process of the tri-layer Al\textsubscript{2}O\textsubscript{3}/TiN/Al\textsubscript{2}O\textsubscript{3} membrane improves the reliability of the conductive layer by sandwiching it between two Al\textsubscript{2}O\textsubscript{3} layers before release. Furthermore, the effect of the film thickness on the transmission (secondary) electron emission will be discussed. The TEY is separated into the transmitted fraction (FSEY) and TSEY, which will be used to characterize the PE interaction in the thin films.   

\newpage

%%%%%%%%%%%%%%%%%% Experimental %%%%%%%%%%%%%%%%%%

\section{Materials \& methods}
\label{sec:exp}
\subsection{Preparation of samples}
\label{ssec:sample}

\begin{figure}[b!]
\centering
\includegraphics[width=1\textwidth,origin=c,angle=0]{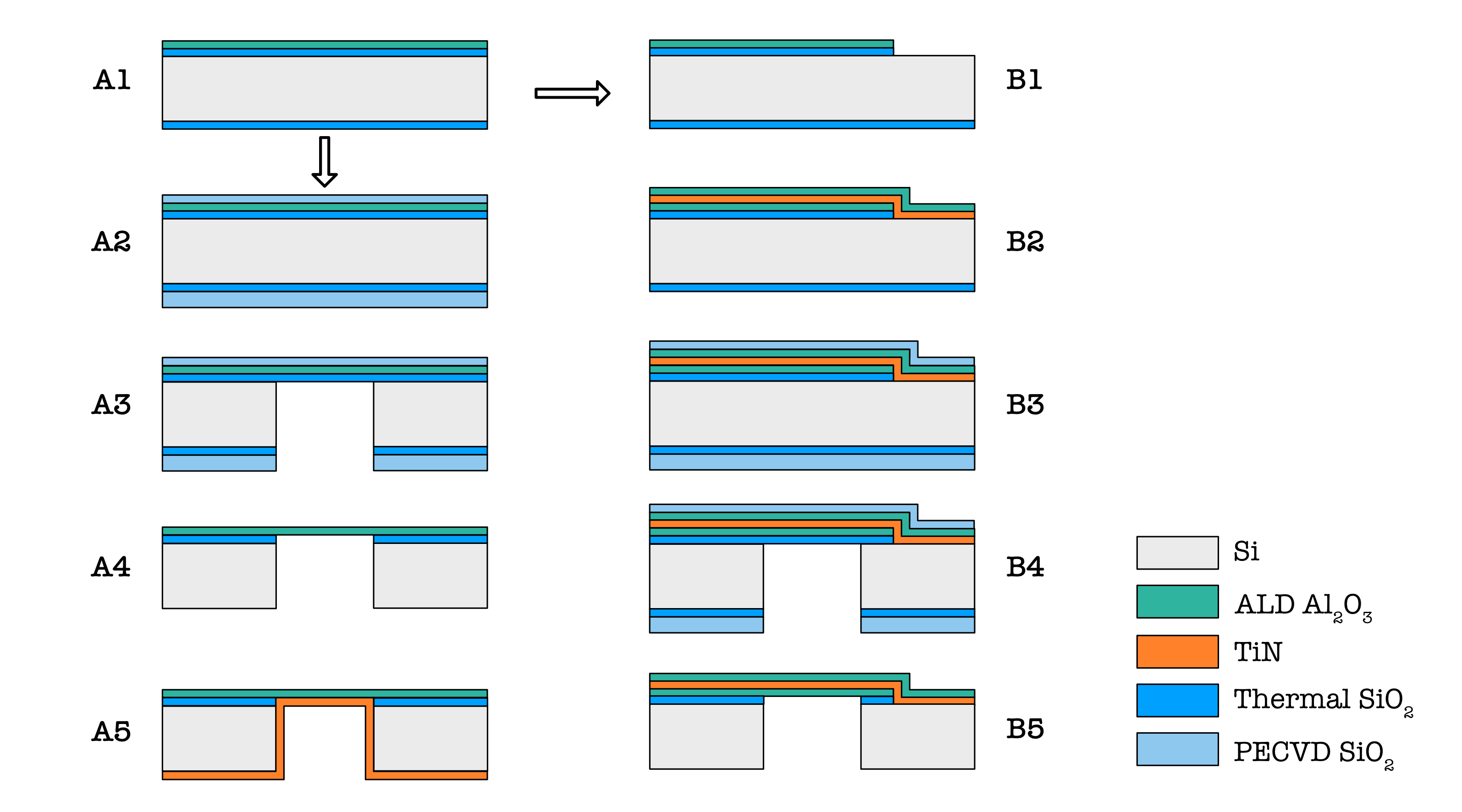}
\caption{\label{fig:3} (A1-A5) Flow chart of the fabrication process of the TiN/Al\textsubscript{2}O\textsubscript{3} Bi-layer membrane. (B1-B5) Flow chart of the Al\textsubscript{2}O\textsubscript{3}/TiN/Al\textsubscript{2}O\textsubscript{3} Tri-layer membrane.}
\end{figure}

The fabrication process of the ultra-thin composite membranes is similar to the fabrication process of tynodes presented in ref. \cite{Prodanovic2018}, but the process is simplified by omitting the support mesh. Instead, a single square membrane with a width of \SI{400}{\micro\meter} is released from the substrate. This basic design is not intended to be used in an actual detector, but is designed with the goal to characterize the transmission secondary electron emission of the multi-layer membranes. In figure \ref{fig:3}, the flowcharts of the fabrication process of two types are given: a TiN/Al\textsubscript{2}O\textsubscript{3}  bi-layer and a Al\textsubscript{2}O\textsubscript{3}/TiN/Al\textsubscript{2}O\textsubscript{3}  tri-layer membrane. The conductive layer is applied as a post-process in the former (figure \ref{fig:3} A5 ), while it is integrated in the process flow of the latter (figure \ref{fig:3} B2). The additional alumina layer serves as a protection layer against the hydrofluoric (HF) vapor etch (figure \ref{fig:3} B2).

For the TiN/Al\textsubscript{2}O\textsubscript{3}  bi-layer membrane, a 4-inch p-type (5-10 \SI{}{\ohm\centi\meter}) wafer with a thickness of \SI[separate-uncertainty = true]{500\pm15}{\micro\meter} is used as a substrate. The Si substrate is oxidized in a wet thermal environment at \SI{1000}{\celsius} until \SI{300}{\nano\meter} of silicon dioxide is formed. This layer will act as a stopping layer and as a sacrificial layer in the process. ALD alumina is grown on top in a thermal ALD ASM F-120 reactor using trimethyl-aluminum and water as a precursor and reactant, respectively (figure \ref{fig:3} A1), at a temperature of \SI{300}{\celsius}. The thickness is varied by choosing different numbers of cycles. Plasma-Enhanced Chemical Vapor Deposition (PECVD) silicon dioxide is then deposited on the front side to protect the alumina layer and on the backside as a masking layer (A2). The silicon substrate is removed by Deep-reactive Ion Etching (DRIE) (A3). After this step, the wafer is cleaved into 15-by-15-mm pieces along predefined break lines. For the final release, the silicon dioxide layers are removed in an HF vapor etch chamber (A4). As a last step, titanium nitride is sputtered as a post-process (A5). This allows the thickness of the conductive layer to be varied and optimized. The membranes have a surface area of \SI{400}{\micro\meter} by \SI{400}{\micro\meter} and film thicknesses $[d_{\text{TiN}}/d_{Al_2O_3}]$ of $2.5/\SI{10}{\nm}$, $5.7/\SI{25}{\nm}$ or $5.7/\SI{50}{\nm}$.

For the Al\textsubscript{2}O\textsubscript{3}/TiN/Al\textsubscript{2}O\textsubscript{3}  tri-layer membranes, the process is the same till the first ALD alumina deposition (A1). After this step, a small patch of alumina and silicon dioxide is removed by plasma etching to expose the silicon substrate (B1). Titanium nitride is then sputtered onto the wafer forming a continuous layer that is in contact with the silicon substrate. Another ALD alumina layer is used to encapsulate this layer (B2). This encapsulation is needed to protect the TiN layer against HF vapor in the last step. The next steps are similar to the previous process. PECVD silicon dioxide is applied as protection and masking layer (B3). The silicon is removed by DRIE (B4) and the wafer is cleaved into 15-by-15-mm dies. The membrane is released by HF vapor etching (B5). The membranes have a surface area of \SI{400}{\micro\meter} by \SI{400}{\micro\meter} and film thicknesses $[d_{Al_2O_3}/d_{\text{TiN}}/d_{Al_2O_3}]$ of $5/2.5/\SI{5}{\nm}$ or $12.5/5.7/ \SI{12.5}{\nm}$. 

\subsection{Experimental method}
\label{ssec:method}

The experimental setup is designed to be mounted onto the moving stage of a Thermo Fisher NovaNanoLab 650 Dual Beam SEM. A teflon holder is attached to the stage in which the setup is fixed. Teflon insulates the sample holder electrically from the stage and the chamber. The SEM has an electron source that provides a continuous electron beam with energy ranging from \SI{0.3}{\keV} to \SI{30}{\keV}. The typical beam current for these experiments is in the \SI{}{\pico\ampere} range, but it can be increased to a few \SI{}{\nano\ampere} if necessary. Though, the current is usually kept to a minimum in order to avoid charge-up effects. The operational vacuum level ranges from \SI{1e-5} down to \SI{1e-6}{\milli\bar}. 

\begin{figure}[t!]
\centering
\includegraphics[width=0.7\textwidth,origin=c,angle=0]{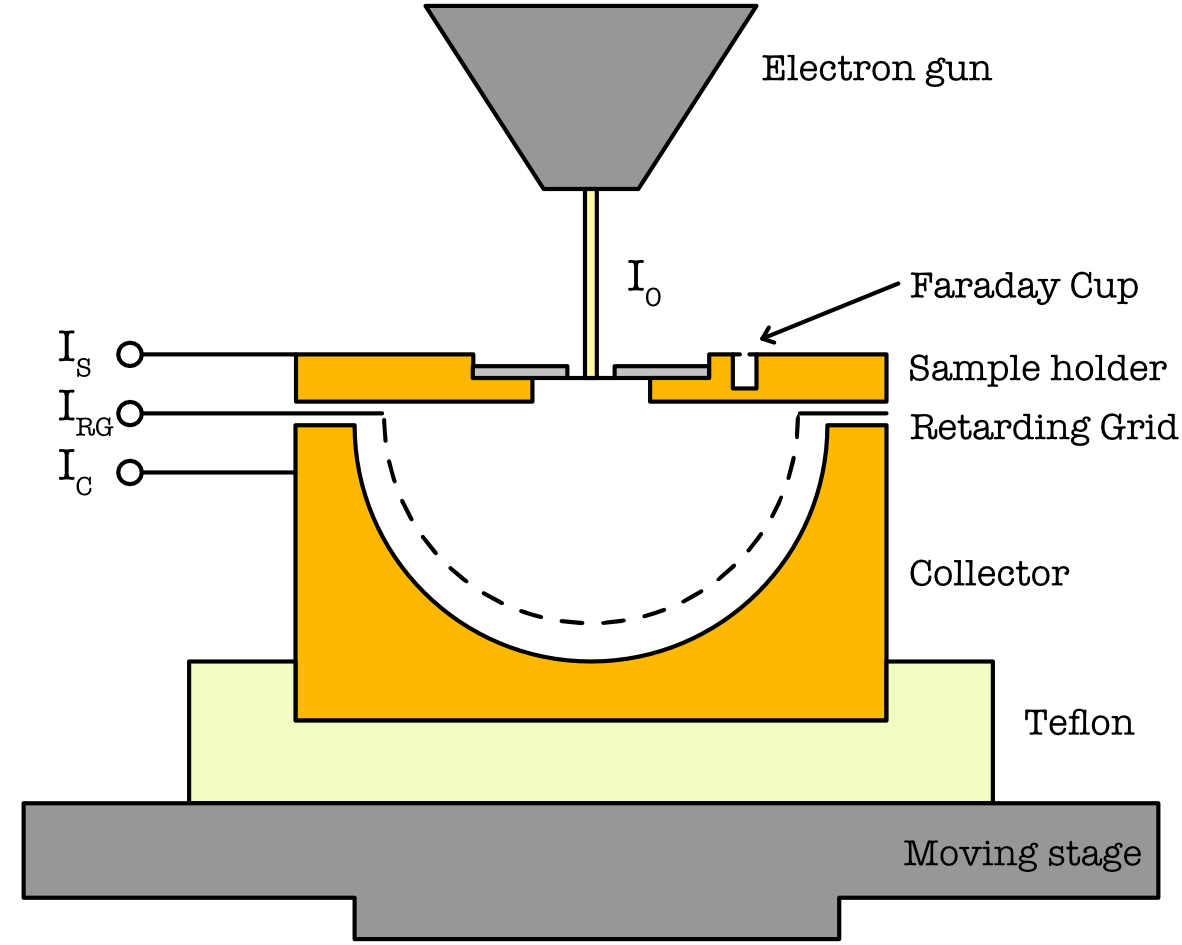}
\caption{\label{fig:4} Schematic drawing of the experimental setup. The primary beam current $I_0$ is measured within the Faraday cup. The sample holder, retarding grid and collector are electrically insulated from each other with Kapton foils. Each is connected via feedhrough to source meters.}
\end{figure}

A schematic representation of the experimental setup is given in figure \ref{fig:4}. It consists of 3 separate electrodes: a collector, retarding grid and sample holder. They are electrically insulated from each other with sheets of kapton foil. Each electrode is connected via a feedthrough to a Keithley 2450 Source meter. This allows each electrode to be biased from \SI{-200}{\volt} to \SI{+200}{\volt}, while measuring the currents simultaneously. The sample is clamped inside the copper sample holder. Silver emulsion is applied on the silicon substrate of the samples to ensure good electrical contact between the sample and holder. Prior to the measurements, the primary beam current $I_0$ as a function of the electron beam energy $E_0$ is measured within a Faraday cup, which is drilled into the sample holder close to the sample. The beam current is stable over the course of a day, so measuring the current once for each beam energy is sufficient.

At the start of the measurement, the electron beam is moved towards the 'active' membrane on the sample. The measurement is performed in image acquisition mode, which has the benefit that only the imaged surface is being irradiated by the beam. Corrections are not needed for any induced currents on the surrounding 'inactive' parts of the sample. Also, charge-up effects are mitigated by distributing the beam over a larger surface. The continuous surface scan has a horizontal field width of \SI{366}{\micro\meter} and a vertical field width of \SI{316}{\micro\meter} with a resolution of 1024 x 884 pixels. This is approximately \SI{0.116}{\square\milli\meter} over which the current is spread. The dwell time (per pixel) is \SI{1}{\micro\second} and the frame time is \SI{0.94}{\second}. For each beam energy $E_0$, the surface is scanned for \SI{20}{\s}, in which multiple frames are taken, before shifting to a higher energy. The background current is measured before and after each reading. By taking multiple frames, charge-up effects can be identified with the SEM: the contrast of the image will change in case of charging. Also, the emission currents will change over time as well. If the in-plane conductivity of the sample is sufficient, the emission currents remain constant. In this case, the average emission current is used to determine the yields. 
\\

This method is a combination of a sample-biasing and a classical collector method; the transmission current is measured directly in the collector, while the reflection current is determined indirectly by subtracting the transmission current from the sample current. The method distinguishes fast electrons $(E_\text{se} > \SI{50}{\eV})$ from true secondary electrons $(E_\text{se} < \SI{50}{\eV})$ by biasing the electrodes in the measurement setup. This requires two separate measurements where the sample is first negatively biased (\SI{-50}{\volt}) and then positively biased (\SI{+50}{\volt}). The primary electron energy $E_0$ is modified by $-\SI{50}{\eV}$ for the former and $+\SI{50}{\eV}$ for the latter.

For a negative bias, the sample holder, retarding grid and collector are biased to \SI{-50}{\volt}, \SI{0}{\volt} and \SI{0}{\volt}, respectively. The negative bias repels fast and slow electrons from the sample on the reflection and transmission side. The transmission coefficient $\sigma_T(E_0)$ is determined by measuring the transmission current with the retarding grid and collector and is given by:
\begin{equation}
\label{eq:7}
\sigma_T(E_0) = \frac{I_{RG-}+I_{C-}}{I_0}
\end{equation}
where $E_0$ is the electron energy of the primary electron, $I_0$ is the primary beam current, $I_{RG-}$ is the retarding grid current and $I_{C-}$ collector current. The minus-subscript indicates that the current is measured under a negative sample bias. The total emission $\sigma(E_0)$, which is the sum of the reflection coefficient $\sigma_R(E_0)$ and transmission coefficient $\sigma_T(E_0)$, is determined by measuring the sample current and is given by:

\begin{equation}
\label{eq:8}
\sigma(E_0) = \frac{I_0-I_{S-}}{I_0}
\end{equation}
where $I_{S-}$ is the sample current. The reflection coefficient is then given by:
\begin{equation}
\label{eq:9}
\sigma_R(E_0) = \sigma(E_0) - \sigma_T(E_0) = \frac{I_0-I_{S-}-I_{RG-}-I_{C-}}{I_0}
\end{equation}

An additional measurement with a positive biased sample can be performed to separate the fast electrons from the slow ones. The sample holder, retarding grid and collector are biased to \SI{+50}{\volt}, \SI{0}{\volt} and \SI{0}{\volt}, respectively. The positive voltage retracts the slow electrons to the sample, while allowing fast electrons $(E_\text{se} > \SI{50}{\eV})$ to escape. The retarding grid prevents tertiary electrons from the collector wall (i.e. unwanted SEs induced within the setup) to flow back towards the sample. The FSE coefficient $\eta_{T}(E_0)$ is determined by measuring the transmission current with the retarding grid and collector and is given by:
\begin{equation}
\label{eq:10}
\eta_T(E_0) =  \frac{I_{RG+}+I_{C+}}{I_0}
\end{equation}
where $E_0$ is the electron energy of the primary electron, $I_0$ is the primary beam current, $I_{RG+}$ is the retarding grid current and $I_{C+}$ collector current, where the plus-subscript indicates a positively biased sample.  Since $\sigma_T(E_0)= \eta_T(E_0)+ \delta_T(E_0)$, the TSE coefficient $\delta_T(E_0)$ is given by
\begin{equation}
\label{eq:11}
\delta_T(E_0) =  \frac{I_{RG-}-I_{C-}}{I_0}-\frac{I_{RG+}+I_{C+}}{I_0}
\end{equation}
The sample current $I_{S+}$ is again the sum of the reflection and transmission current. In this case
\begin{equation}
\label{eq:12}
\eta_T(E_0) + \eta_R(E_0)=  \frac{I_0-I_{S+}}{I_0}
\end{equation}
After substituting $\eta_T(E_0)$, the BSE coefficient $\eta_R(E_0)$ is given by:
\begin{equation}
\label{eq:13}
\eta_R(E_0) =  \frac{I_0-I_{S+}-I_{RG+}-I_{C+}}{I_0}
\end{equation}
The RSE coefficient $\delta_R(E_0)$ can be determined by using the definition of the total emission coefficient: $\sigma(E_0) = \eta_R(E_0) + \delta_R(E_0) + \eta_T(E_0) + \delta_T(E_0)$, from which it follows that
\begin{subequations}\label{eq:14}
\begin{align}
\label{eq:14:1}
\delta_R(E_0) &=  \sigma(E_0) - \eta_T(E_0) - \delta_T(E_0) - \eta_R(E_0) 
\\
\label{eq:14:2}
\delta_R(E_0) &=  \frac{I_0-I_{S-}-I_{RG-}-I_{C-}}{I_0} -  \frac{I_0-I_{S+}-I_{RG+}-I_{C+}}{I_0}
\end{align}
\end{subequations}
With eq. \eqref{eq:11}, \eqref{eq:12}, \eqref{eq:13} and \eqref{eq:14:2}, all relevant yields can be calculated from the measured currents.  

%%%%%%%%%%%%%%%%%% Results %%%%%%%%%%%%%%%%%%

%\newpage

\section{Results}
\label{sec:results}
\subsection{Bi-layer membrane}
\label{ssec:bi}

\begin{figure}[b!]
\centering 
\includegraphics[width=0.8\textwidth,origin=c,angle=0]{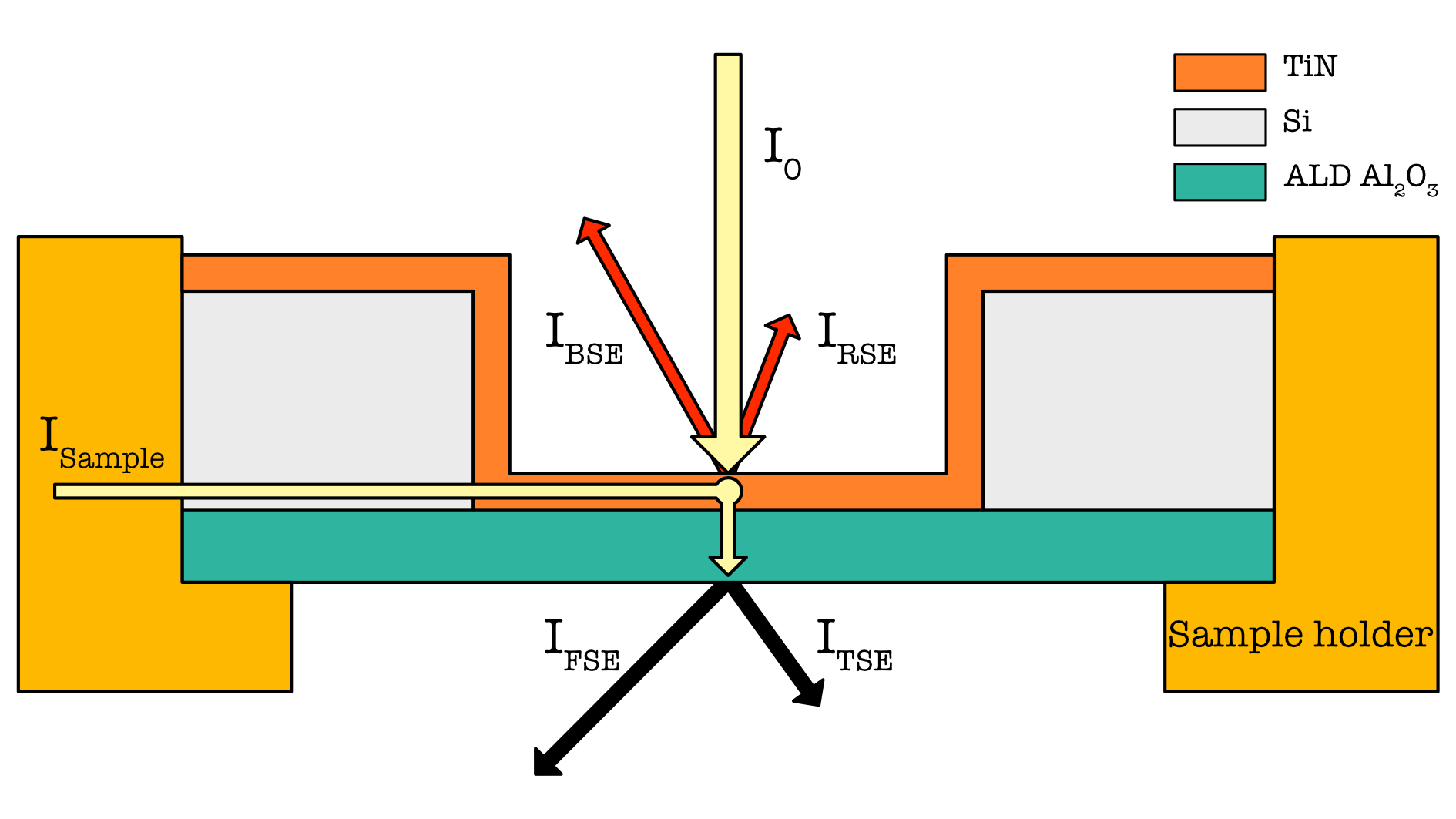}
\caption{\label{fig:5} The currents to and from a bi-layer membrane irradiated by an electron beam. TiN is sputtered into the window opening on the reflection side to provide a conductive path from the sample holder to the irradiated region.}
\end{figure}

In figure \ref{fig:5}, a schematic drawing of a bi-layer membrane is given with all the currents that flow to and from the irradiated region. The flat side of the sample with the ALD alumina layer is facing downwards in the transmission direction, while the window opening in the silicon substrate is facing upwards. The conductive TiN layer is deposited inside the window opening. 

\begin{figure}
  \centering
	\subfloat[2.5 / \SI{10} {\nano\meter}]{\includegraphics[scale=0.135]{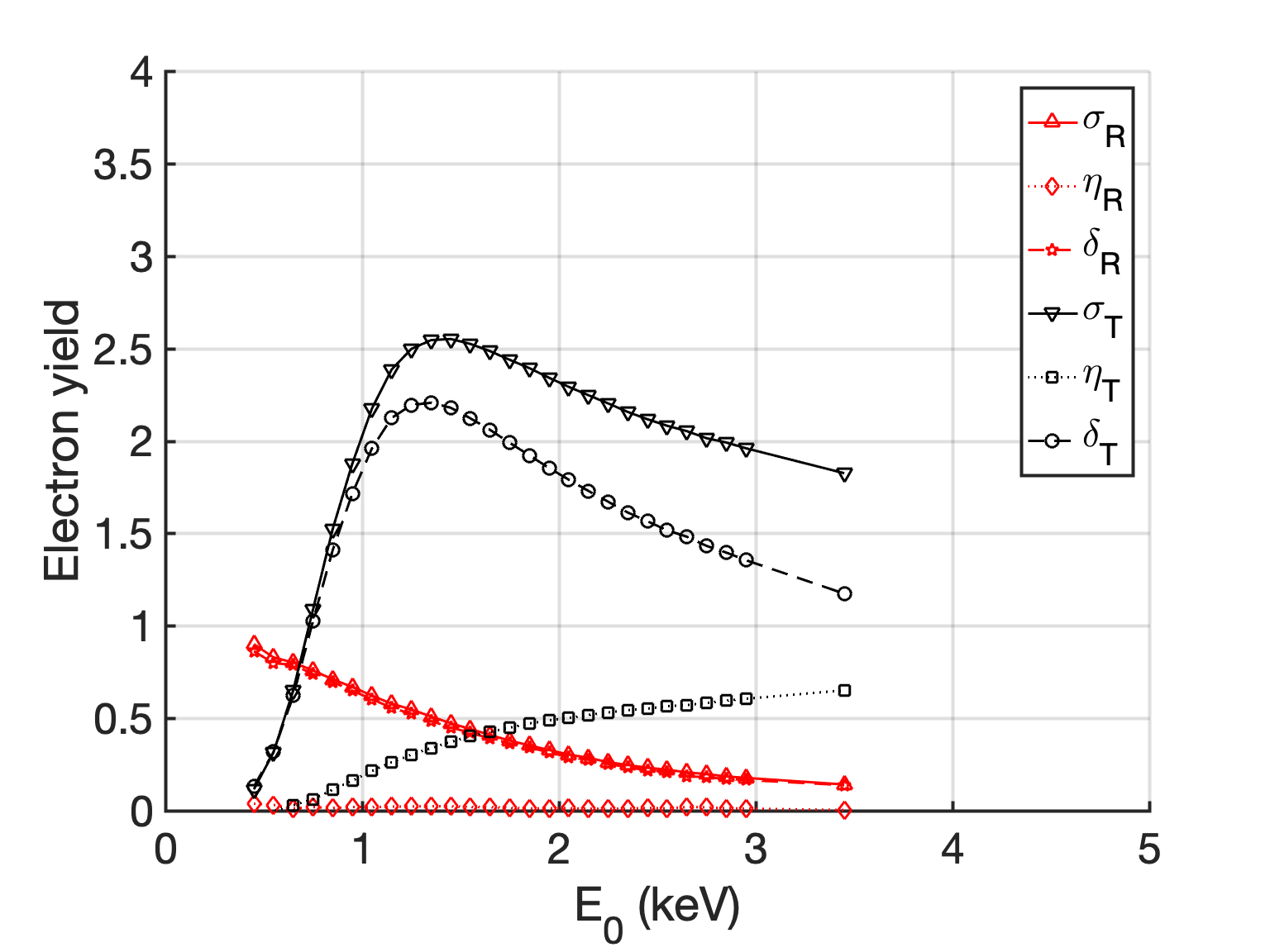}}
	\subfloat[5.7 / \SI{25} {\nano\meter}]{\includegraphics[scale=0.135]{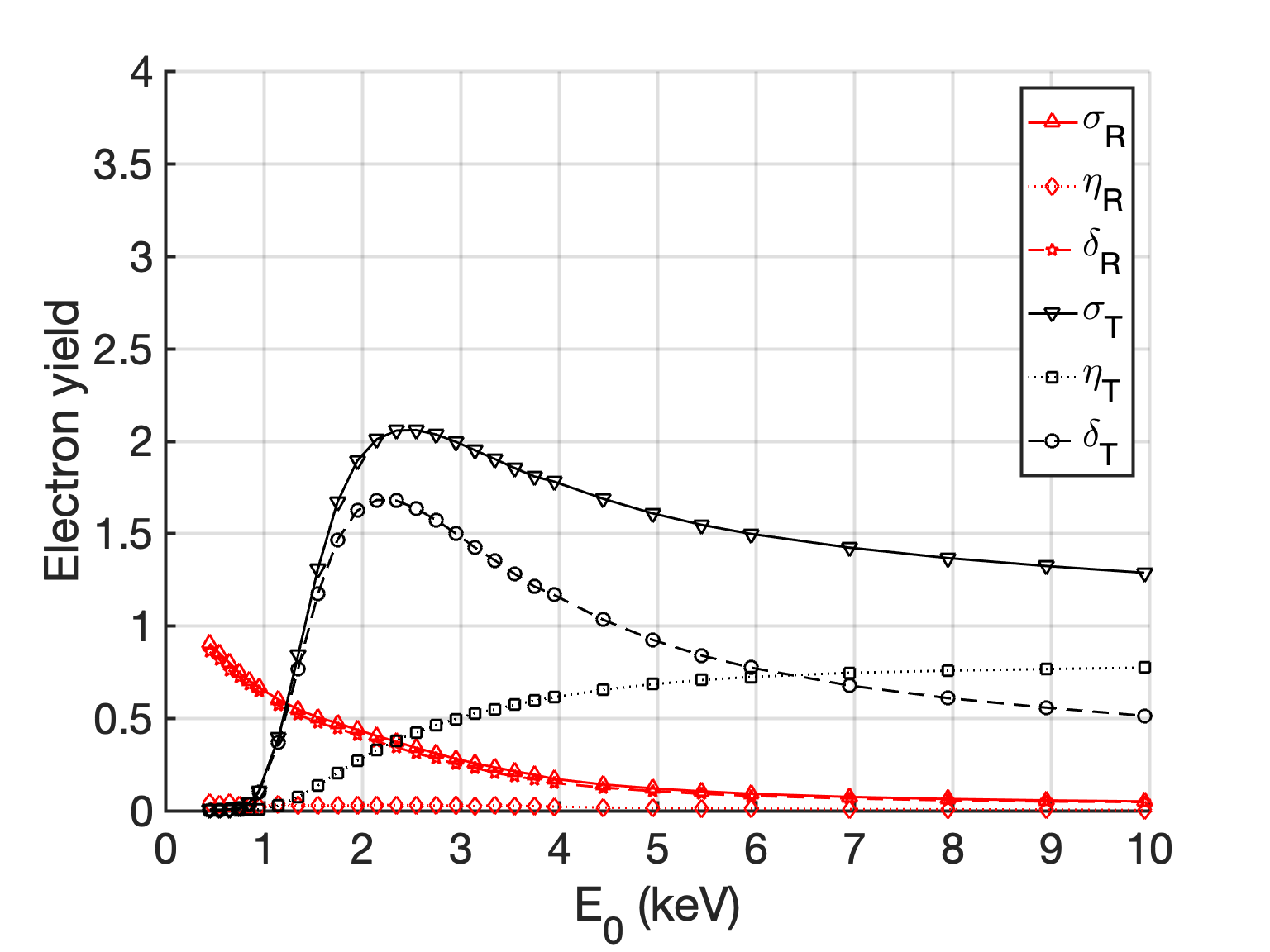}}
	    \quad
	\subfloat[5.7 / \SI{50} {\nano\meter}]{\includegraphics[scale=0.135]{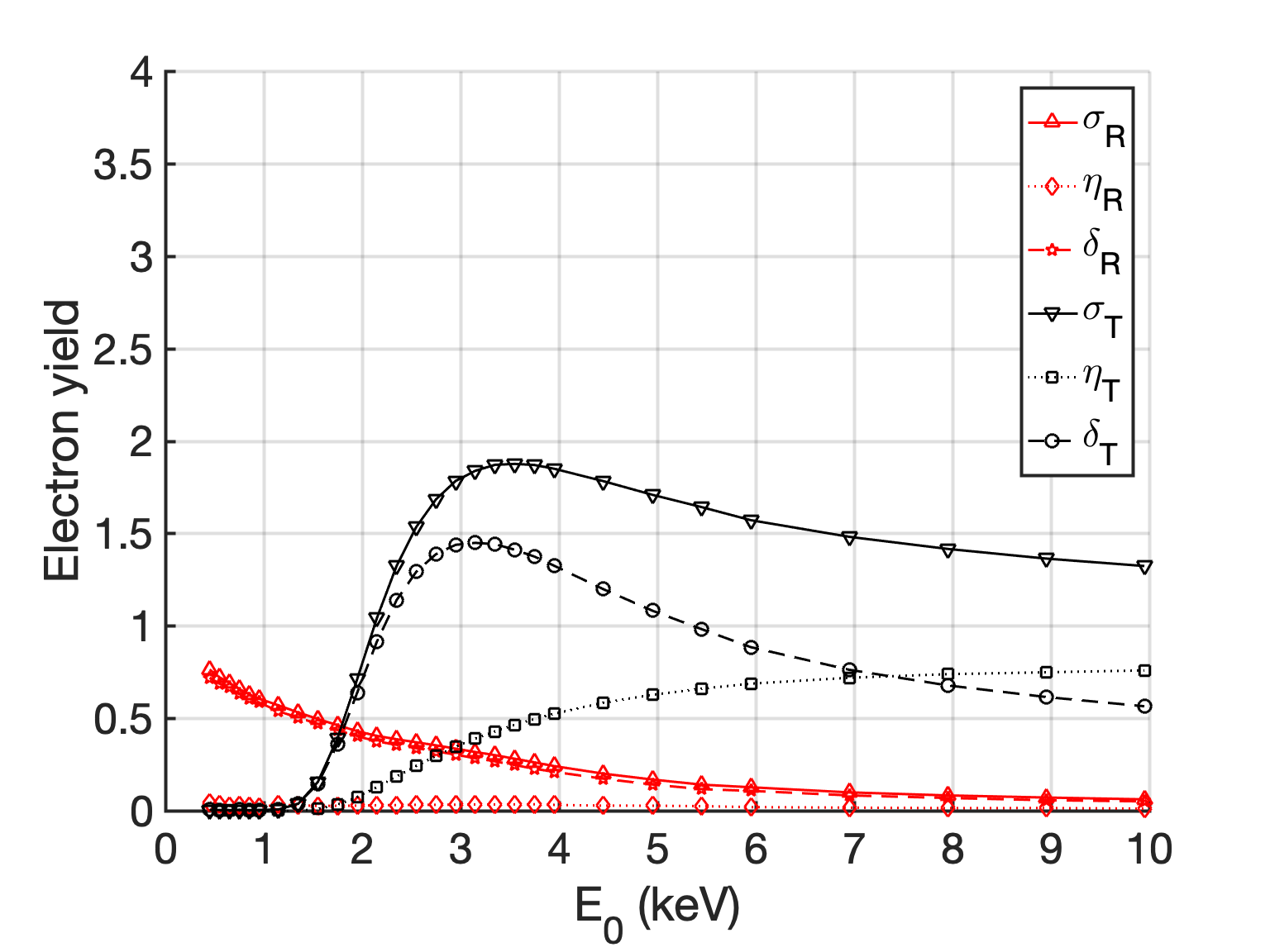}}
     \caption{Electron emission coefficients of a bi-layer membrane TiN/Al\textsubscript{2}O\textsubscript{3} with thicknesses $d_{TiN} / d_{Al_2O_3}$} \label{fig:6}
\end{figure}

	The (secondary) electron yield curves as a function of the primary electron energy $E_0$ are given in figure \ref{fig:6} for bi-layer TiN/Al\textsubscript{2}O\textsubscript{3} membranes with thicknesses of 2.5/10, 5.7/25 and 5.7/50 \SI{}{\nano\meter}, respectively. The total effective film thickness $d$ is calculated with equation \eqref{eq:6}. For low-$Z$ materials, the transmission parameters are assumed to be approximately equal: $p_1\cong p_2$. The conversion factor is then simply the ratio between the ranges: $R_{Al_{2}O_{3}}/R_{\text{TiN}} \cong 1.51$, i.e. the TiN layer can be replaced by an Al\textsubscript{2}O\textsubscript{3} layer with an effective thickness that is 1.51 times larger. It is listed in table \ref{tab:1} along with the transmission yield curve characteristics; the critical energy $E_c$, maximum TEY $\sigma_{T}^\text{max}(E_{T}^{\text{max}})$ and maximum TSEY $\delta_{T}^\text{max}(E_{\text{TSE}}^\text{max})$.
	
	The reflection SEE coefficients are represented by the red curves in figure \ref{fig:5}. For a bi-layer membrane, the contribution to the reflection yields is from the TiN layer. The BSE coefficient $\eta_R(E_0)$ is close to zero for all three thicknesses. There are two factors that contribute to this low value. First, the BSE yield of membranes and foils are expected to be lower in comparison with their bulk counterpart \cite{Kanter1961a,Holzl1969}. Second, the silicon window frame reduces the field of view for BSEs and will recapture some. The RSE coefficient $\delta_R(E_0)$ is below 1 and is lower than expected. The reduction in yield can again be attributed to recapture. The maximum REY on a bulk TiN sample can range from 1.4 to 2.8 for a PE energy of \SI{300}{\eV} depending on the deposition technique and conditions \cite{He2004,Wang2018}. 
	
	The reduction in REY due to the silicon window frame is estimated by using a test sample. In appendix \ref{sec:samplecor}, a p-type silicon membrane with widths of \SI{400}{\micro\meter} is used to estimate the reduction in yield due to recapture. The aspect ratio of the window and the wall is $\sim$1.2 and is the same as the other samples. The emission surface of the silicon membrane on both sides is identical, so the difference in yield is solely due to the presence of the window walls on one side. When the window opening is facing the electron gun, the REY was reduced by 35 to 45\%. When the window opening was facing away, the TEY was reduced by 15 to 30 \%.

	The transmission SEE coefficients are represented by the black curves in figure \ref{fig:6}. The transmission side consists of Al\textsubscript{2}O\textsubscript{3}. The FSE coefficient $\eta_T(E_0)$ is the fraction of the primary electron beam that penetrates through the membrane and contains electrons with energy $E > \SI{50}{eV}$. Thin films become transparent for high-energetic electrons. As such, almost all PEs should be collected by the collector, i.e. the FSE curve approach 1 for high PE energies. However, the curves converge to 0.8 instead. The discrepancy can be attributed to (back)scattering of the transmitted PEs on the retarding grid and the collector wall, which will induce tertiary currents that can lower the net transmission current. The effect of tertiary currents on the transmitted fraction will be discussed in \ref{ssec:fraction}. In appendix \ref{sec:meascor}, a correction term is estimated by taking scattering events in the collector into account. 
	
\begin{table}[t!]
\caption{\label{tab:1} Summary of important electron emission values of all composite membranes. The total effective thickness $d$ is calculated with eq. \eqref{eq:6}. The density of ALD Al\textsubscript{2}O\textsubscript{3} and sputtered TiN are $\SI{3.1}{\g\per\cm^{3}}$ and $\SI{5.2}{\g\per\cm^{3}}$, respectively.} 
\smallskip
\centering
\begin{tabular}{| c | c | c | c | c | c | c | c | c | c | c |}
\hline
Type & $d_{Al_2O_3} $ & $d_{\text{TiN}}$ & $d_{Al_2O_3}$ & $d$ & $d$ & $\sigma_{T}^\text{max}$ & $E_T^\text{max}$ & $\delta_T^\text{max}$ & $E_\text{TSE}^\text{max}$ & $E_c$\\
 & (\SI{}{\nm}) & (\SI{}{\nm}) & (\SI{}{\nm}) & (\SI{}{\nm}) & (\SI{}{\ug\per\cm^{2}}) & & (\SI{}{\keV}) & & (\SI{}{\keV}) & (\SI{}{\keV}) \\
\hline
Bi-layer & - & 2.5 & 10 & 13.8 & 4.40 & 2.6 & 1.45 & 2.2 & 1.35 & 0.5\\
Bi-layer & - & 5.7 & 25 & 33.6 & 10.7 & 2.1 & 2.55 & 1.7 & 2.15 & 1.0\\
Bi-layer & - & 5.7 & 50 & 58.6 & 18.5 & 1.9 & 3.55 & 1.5 & 3.15 & 1.4\\
Tri-layer & 5 & 2.5 & 5 & 13.8 & 4.40 & 3.1 & 1.55 & - & - & -\\
Tri-layer & 12.5 & 5.7 & 12.5 & 33.6 & 10.7 & 2.7 & 2.75 & - & - & -\\
\hline
\end{tabular}
\end{table}

	The TSE coefficient $\delta_T(E_0)$ represents electrons with $E_\text{se} < \SI{50}{\eV}$, which originates from the Al\textsubscript{2}O\textsubscript{3} layer within the escape depth. The initial rise of the TSEY curve starts at the threshold energy $E_\text{th}$. At this energy, the first (slow) secondary electrons emerge from the membrane in transmission. It is correlated to the critical energy $E_c$ for which 1\% of the PEs are transmitted. Another characteristic is the maximum TSEY obtained with PEs with energy $E_\text{TSE}^\text{max}$ : $\delta_T^\text{max}(E_\text{TSE}^\text{max})$. The thinnest membrane with a total effective thickness of \SI{13.8}{\nm} has the highest maximum TSEY of 2.21 (\SI{1.35}{\keV}). 
		
	The total transmission coefficient $\sigma_T(E_0)$ is the sum of $\delta_T(E_0)$ and $\eta_T(E_0)$. In literature, the distinction between $\delta_T(E_0)$ and $\eta_T(E_0)$ is often not made. Unless specified, usually the total transmission yield $\sigma_T(E_0)$ is given. The performance of a tynode can be expressed by the maximum TEY: $\sigma_T^\text{max}(E_T^\text{max})$. The highest maximum TEY of 2.55 (\SI{1.45}{\keV}) was measured on a membrane with $d=\SI{13.8}{\nm}$. The maximum TEY and TSEY of the other membranes are listed in table \ref{tab:1}. 

%%%%%%%%%%%%%%%%%%%%%%%%

\subsection{Tri-layer membrane}
\label{ssec:tri}

\begin{figure}[b!]
\centering 
\includegraphics[width=0.8\textwidth,origin=c,angle=0]{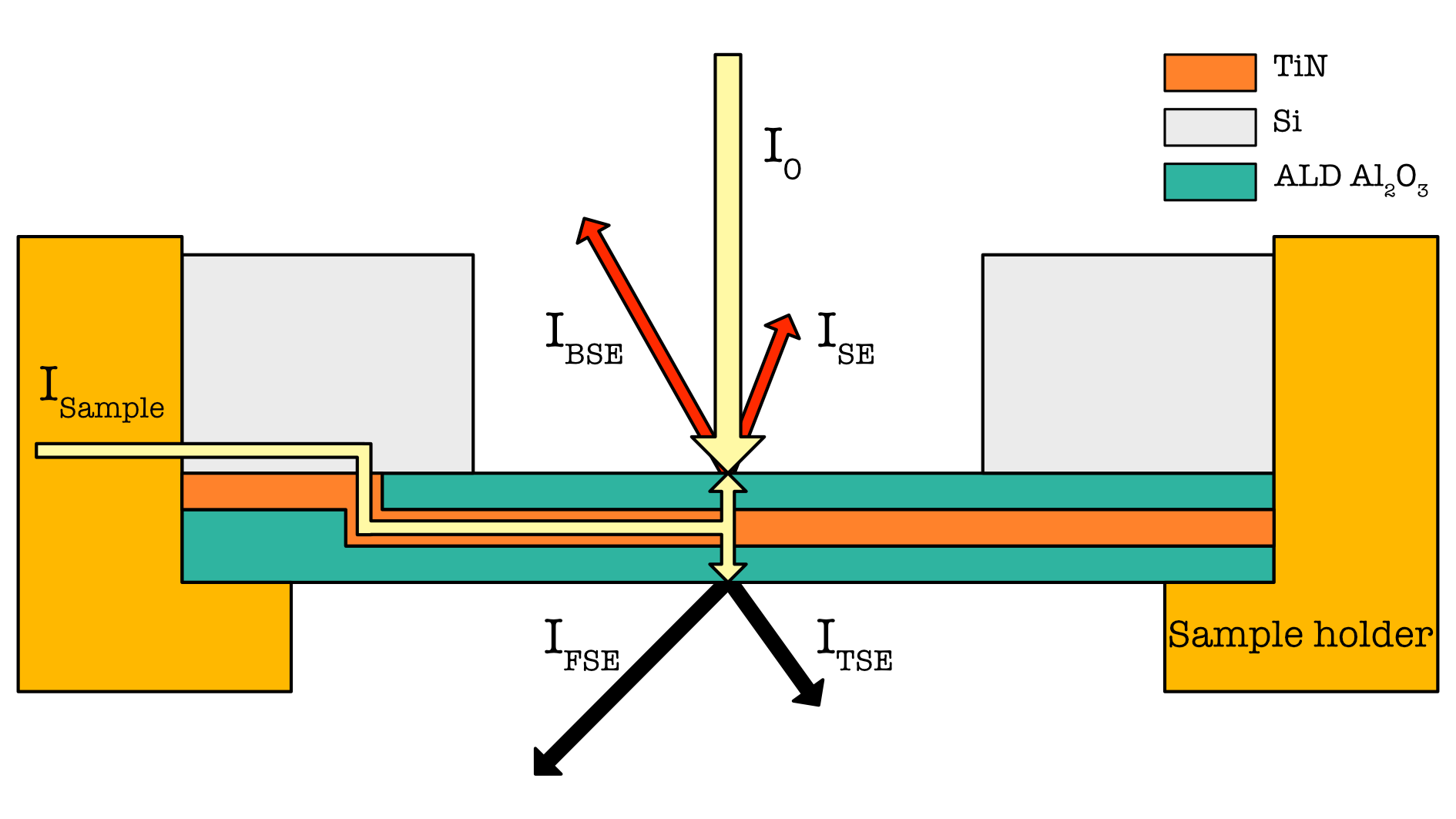}
\caption{\label{fig:7} The currents to and from a tynode with a sandwiched TiN layer irradiated by an electron beam. The reflection side is also covered with Al\textsubscript{2}O\textsubscript{3}, which protects the conductive layer during the fabrication process.}
\end{figure}

	In figure \ref{fig:7}, a schematic drawing of a tri-layer membrane is shown. The TiN layer is sandwiched between two layers of alumina. The three layers are deposited subsequently in the fabrication process, which improves the reliability of the conductive layer. The currents flowing to and from the irradiated regions are indicated by the arrows. 
	
\begin{figure}[b!]
  \centering
	\subfloat[\label{fig:8a}] {\includegraphics[scale=0.135]{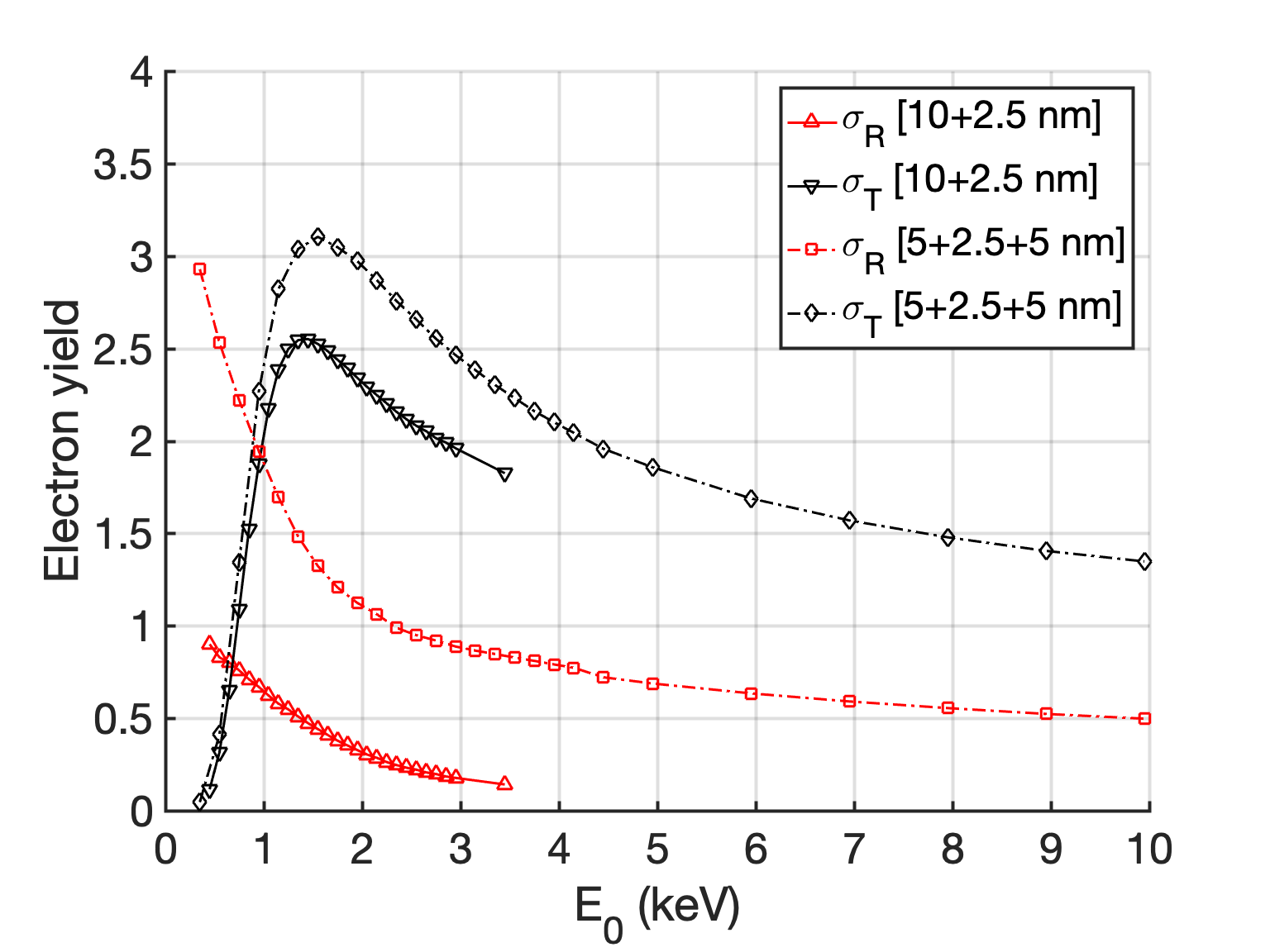}}
	\subfloat[\label{fig:8b}]{\includegraphics[scale=0.135]{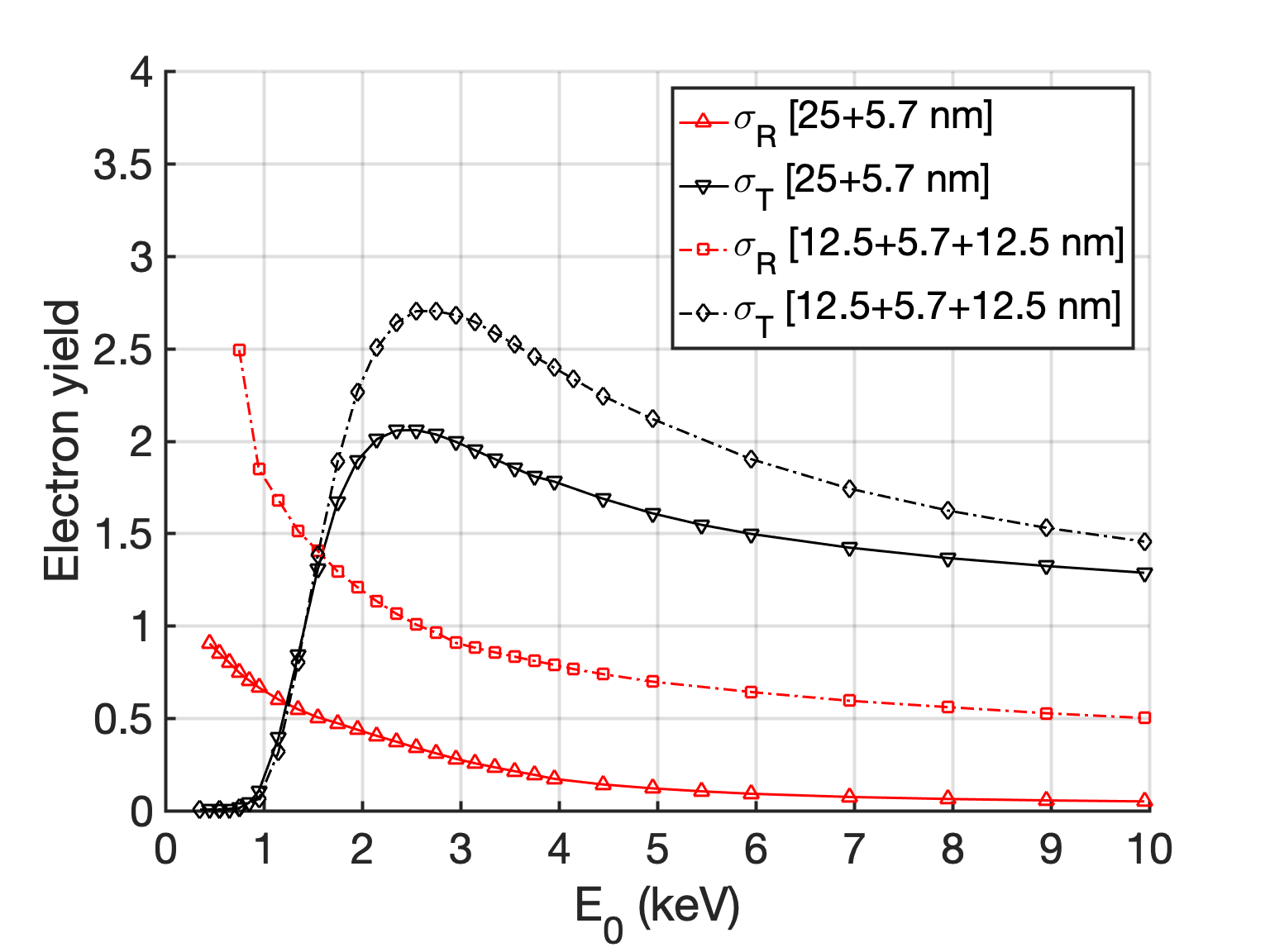}}
     \caption{\label{fig:8} Electron emission yield curves of a bi-layer membrane compared to a tri-layer membrane Al\textsubscript{2}O\textsubscript{3}/TiN/Al\textsubscript{2}O\textsubscript{3} with the same total effective thickness: \protect\subref{fig:8a} $d=\SI{13.8}{\nm}$ \protect\subref{fig:8b} $d=\SI{33.6}{\nm}$}
\end{figure}

	In figure \ref{fig:8a}, the reflection $\sigma_R(E_0)$ and transmission $\sigma_T(E_0)$ coefficients of a bi-layer membrane TiN/Al\textsubscript{2}O\textsubscript{3} are compared to a tri-layer membrane Al\textsubscript{2}O\textsubscript{3}/TiN/Al\textsubscript{2}O\textsubscript{3}. The thicknesses of the layers for the two membranes are 2.5/\SI{10}{\nm} and 5/2.5/\SI{5}{\nm}, respectively, with a total effective thickness of \SI{13.8}{\nm} for both.

	The reflection coefficient $\sigma_R(E_0)$ is significantly smaller for the bi-layer compared to the tri-layer, since the material of the emission surfaces are different. The REY of TiN is lower than that for Al\textsubscript{2}O\textsubscript{3}. Therefore, a direct comparison of the REY is not useful.
	
	The transmission coefficient $\sigma_T(E_0)$ for both type of membrane has the same threshold energy $E_\text{th}$. This shows that both membranes have a similar thickness and stopping power. However, the maximum TEY of $3.1$ (\SI{1.55}{\keV}) is higher for the tri-layer membrane compared to the bi-layer yield of 2.6 (\SI{1.45}{keV}). The better performance is also observed for the membrane with $d=\SI{33.6}{\nm}$ as shown in figure \ref{fig:8b}. The maximum TEY is 2.7 (\SI{2.75}{\keV}) for the tri-layer and 2.1 (\SI{2.55}{\keV}) for the bi-layer. Hence, encapsulating the conductive layer of TiN between two layers of Al\textsubscript{2}O\textsubscript{3} improves the TEY in comparison with the bi-layer membrane.
	
%\newpage

%%%%%%%%%%%%%%%%%% Discussion %%%%%%%%%%%%%%%%%%

\section{Discussion}
\label{ssec:disc}

\subsection{Reflection vs. Transmission yield}
\label{ssec:versus}

\begin{figure}[b!]
\centering 
\includegraphics[width=0.7\textwidth,origin=c,angle=0]{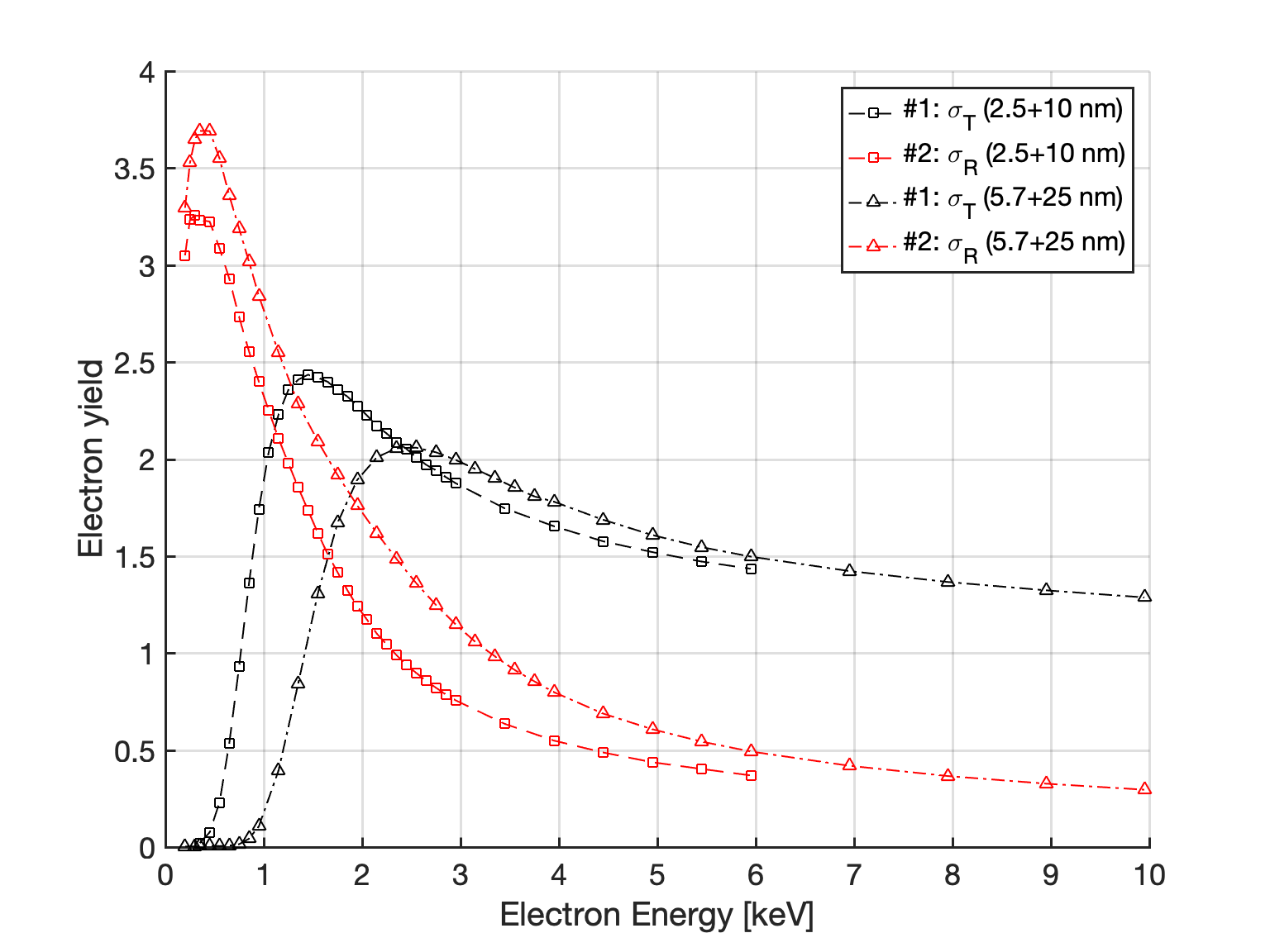}
\caption{\label{fig:9} The combined results obtained from two separate measurements on the same bi-layer sample. In the first measurement, the flat surface is facing downwards to obtain $\sigma_{T}$. In the second measurement, the flat surface is facing upwards to obtain $\sigma_{R}$. The combined results portray the electron emission characteristic of a flat Al\textsubscript{2}O\textsubscript{3} membrane with a thickness of 13.8 nm and one with 33.6 nm.}
\end{figure}

	The REY of a thin film can provide valuable information for the design process of tynodes. Film-on-bulk samples are less complex to manufacture, but can provide insight on the TEY if the ratio between REY and TEY is known. Furthermore, the effect of surface and/or thermal treatment on the REY can be measured on film-on-bulk samples. An estimate can then be made on how the treatment affects the TEY of tynodes. 

	An ideal sample would be a large freestanding symmetrical film without any obstructions on both side. Regrettably, such sample cannot be made due to the fragility of ultra-thin films. However, a method to circumvent the shortcoming of our samples is to perform an additional measurement on the bi-layer membranes and combine the results of the two separate measurements. In the first measurement, the Al\textsubscript{2}O\textsubscript{3}-layer is facing downwards, so emission in transmission is unobstructed. In the second measurement, the sample is flipped over so that the Al\textsubscript{2}O\textsubscript{3}-layer is facing upwards. By combining the TEY of the first measurement and the REY of the second, the electron emission characteristic of a flat Al\textsubscript{2}O\textsubscript{3} membrane is represented. The combined yield curves are shown in figure \ref{fig:9} for the bi-layer membranes with $d=\SI{13.8}{\nm}$ and $d=\SI{33.6}{\nm}$. The maximum REY $\sigma_R^\text{max}(E_R^\text{max})$ is 3.3 (\SI{0.30}{\keV}) and 3.7 (\SI{0.35}{\keV}), respectively. This result is close to the maximum REY of an ALD Al\textsubscript{2}O\textsubscript{3}-film (\SI{12.5}{\nm}) on bulk silicon sample, which has a $\sigma_R^\text{max}$ of 3.6 (\SI{0.4}{keV}) \cite{VanderGraaf2017}.

	When comparing the REY curves in figure \ref{fig:9}, the curve of the thinner film with $d=\SI{13.8}{\nm}$ is lower for all energies compared to the film with $d=\SI{33.6}{\nm}$. The RSEY comprises of SEs generated by primary electrons $\delta_p$ and back-scattered (primary) electrons $\delta_b$: $\delta_R=\delta_p+\delta_b$ \cite{Kanaya1972}. In bulk samples (and thick films), a large contribution to RSE generation comes from BSEs that dissipate energy when they return from the interior. In an experiment, where an aluminum target was irradiated with keV-electrons, the back-scattered electrons contributed close to 40\% of the generated RSEs \cite{Kanter1961a}. Also, backscattered electrons were found to be 4.9 times as effective in generating SEs compared to incoming PEs. In thin films, the backscatter contribution $\delta_b$ is negligible when $R(E_0)\gg d$, since most PEs will be transmitted through the film. The lower REY of the thinner film can be attributed to the reduced backscatter contribution $\delta_b$. A similar graph was found for thin Al films and Al bulk material by Kanter \cite{Kanter1961a}. As such, the thickness $d=\SI{13.8}{\nm}$ is near the optimal thickness for Al\textsubscript{2}O\textsubscript{3} films. Reducing the thickness further, the REY will decrease and in ratio also the TEY. 

%%%%%%%%%%%%%%%%%%%%%%%%

\subsection{Transmitted fraction}
\label{ssec:fraction}

\begin{figure}[b!]
  \centering
	\subfloat[\label{fig:10a}] {\includegraphics[scale=0.135]{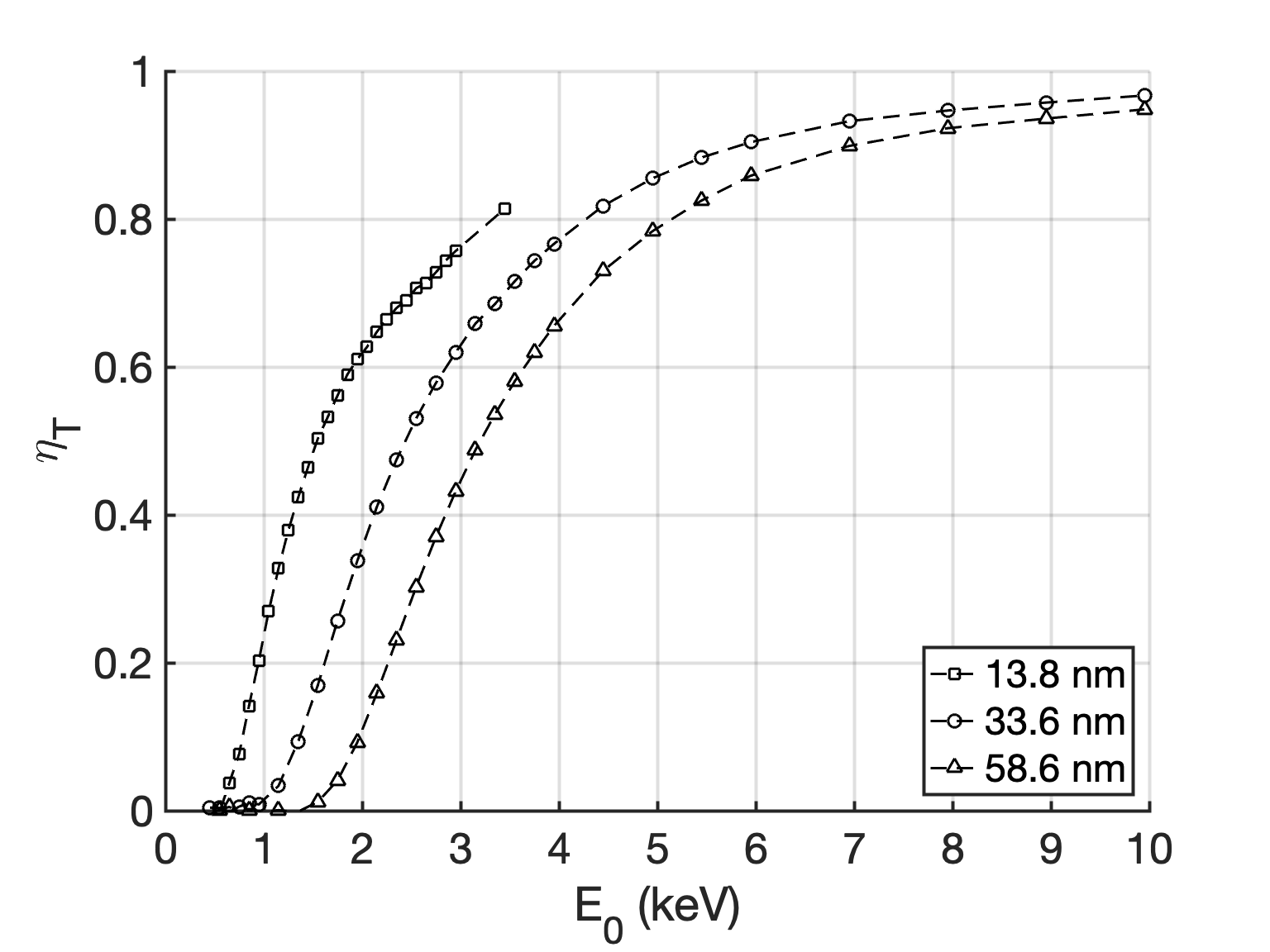}}
	\subfloat[\label{fig:10b}]{\includegraphics[scale=0.135]{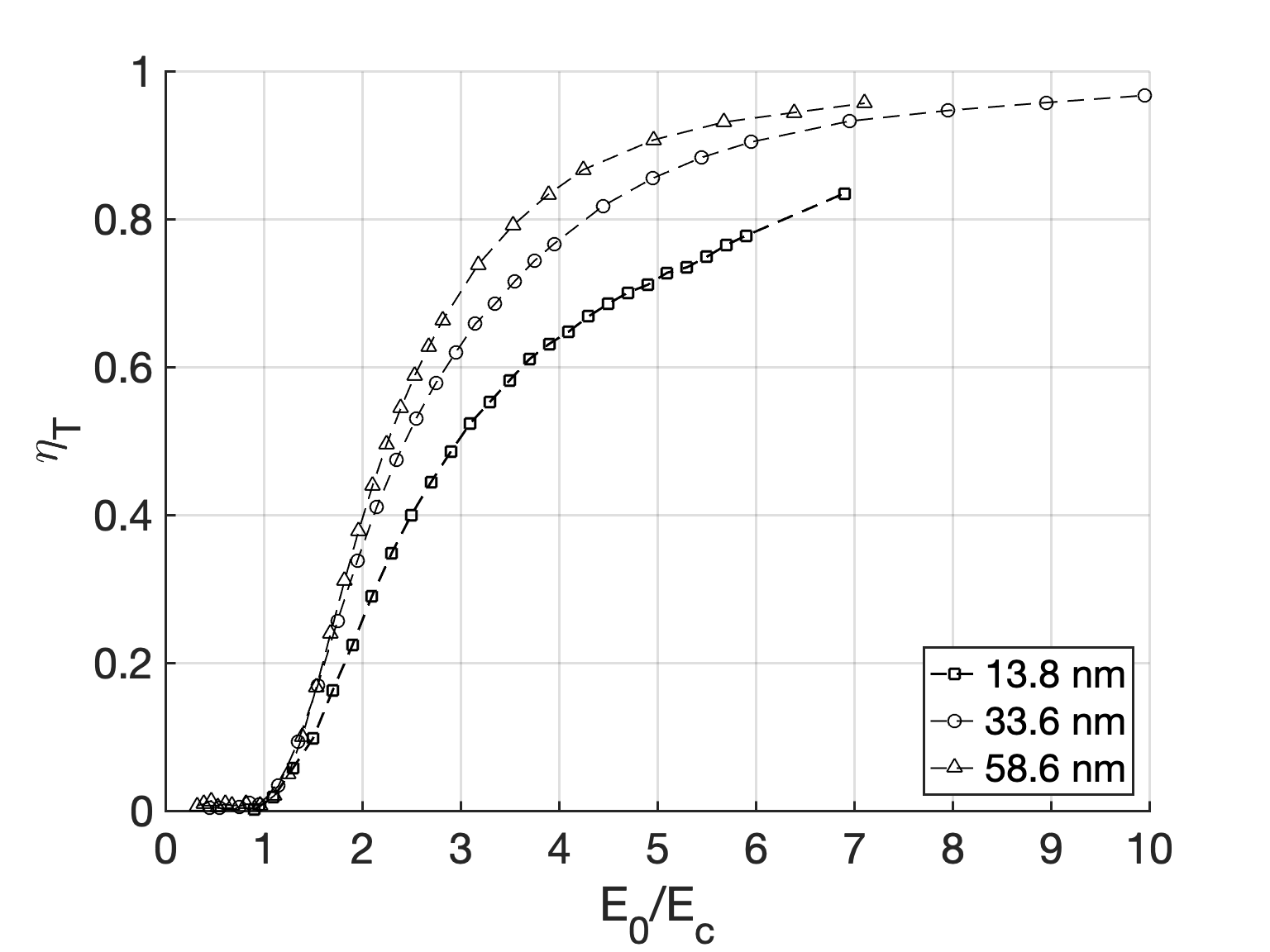}}\\
	\subfloat[\label{fig:10c}]{\includegraphics[scale=0.135]{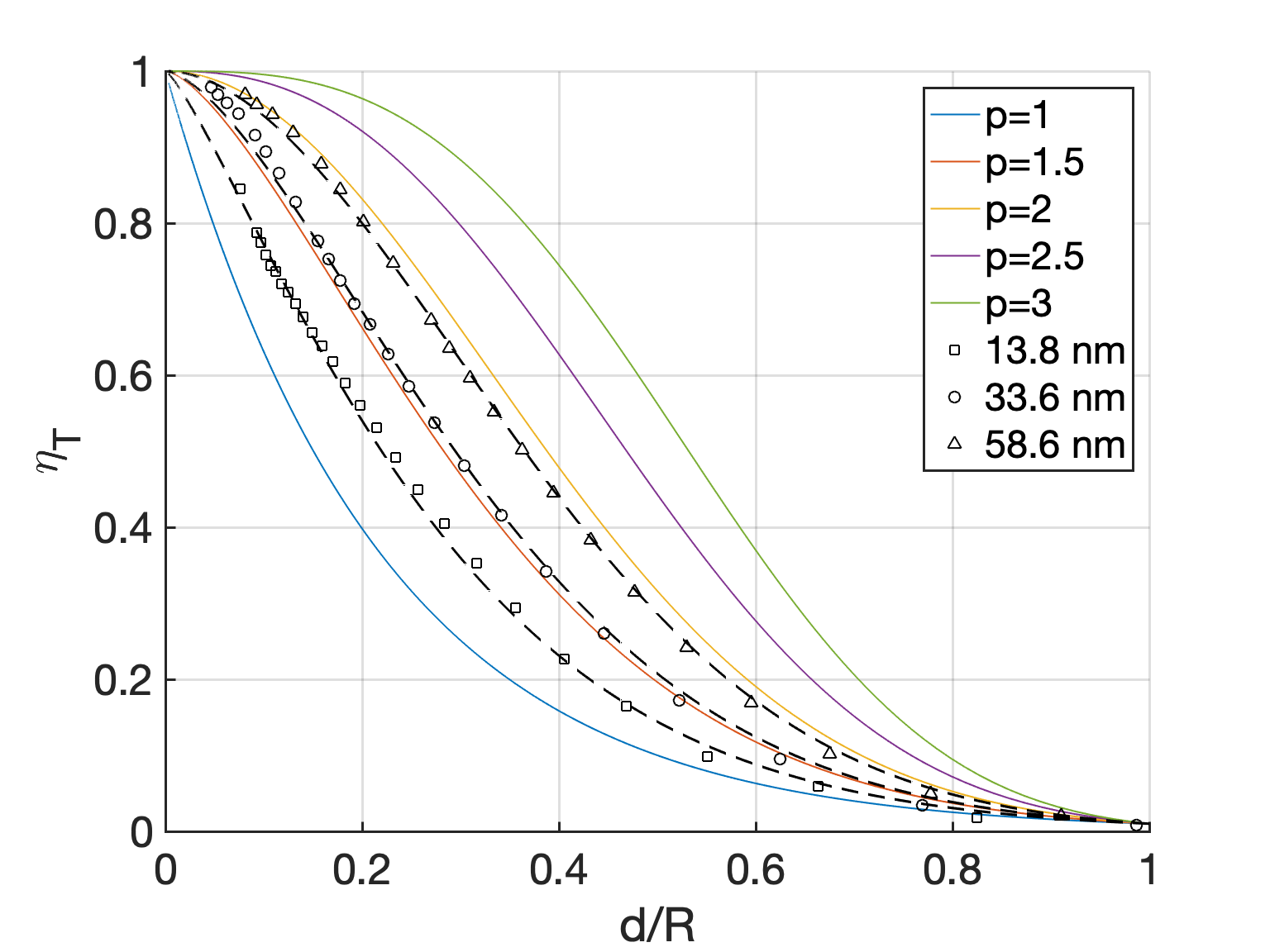}}\\
     \caption{Transmitted fraction as a function of \protect\subref{fig:10a}  the primary electron energy, \protect\subref{fig:10b} the reduced inital energy $E/E_c$ and \protect\subref{fig:10c} the reduced thickness $d/R$ of the membranes. The p-value for the three film thicknesses are 1.25, 1.55 and 1.88, respectively.}
\end{figure}

The FSE coefficient $\eta_T$ is defined as the ratio between the number of initial and transmitted PEs. In early experiments, this ratio is often referred as the transmitted fraction and is used to obtain electron-range relations, such as equation \eqref{eq:5} \cite{Fitting1974}. The transmission characteristic of PEs passing through thick films is material-dependent and well-defined. It can be represented by a universal transmission curve \cite{Kanter1961b} or a (constant) transmission parameter \cite{Fitting1974}. However, for ultra-thin films the transmission characteristics deviate. 

In figure \ref{fig:10a}, the transmitted fraction of the bi-layer films with different thicknesses is shown. A correction term of 0.2 has been applied to $\eta_T(E_0)$ to account for tertiary currents in the semi-spherical collector (see appendix \ref{sec:meascor}). In figure \ref{fig:10b}, the transmitted fraction is plotted against the reduced initial energy $E_0/E_c$. This normalization was proposed by Kanter to define the transmission characteristics of electrons through foils of various materials with different thicknesses \cite{Kanter1961b}. According to Kanter, the transmission curve approaches a (material-dependent) universal transmission curve for large film thickness ($d_c \geq \SI{20}{\micro\gram/\square\cm}$). However, for thinner films the curve deviated as was shown for carbon foils. The same deviation is also observed in figure \ref{fig:10b} for the films with $d=\SI{13.8}{\nm}$ and $d=\SI{33.6}{\nm}$, while the curve of the thicker film seems to converge towards a universal curve. For ALD Al\textsubscript{2}O\textsubscript{3} with a density of $\SI{3.1}{\g\per\cm^{3}}$, a universal curve is expected to be found for a film thickness of $d_c \geq \SI{64.5}{nm}$. This is in agreement with the results of Kanter.  

A different normalization was proposed by Fitting \cite{Fitting1974} in which the transmitted fraction is expressed as a function of the reduced film thickness $d/R$ as shown in figure \ref{fig:10c}. The transmitted fraction can then be expressed by: 
\begin{equation}
\label{eq:15}
\eta_T(E_0) =  \exp \left[-4.605 \left(\frac{d}{R(E_0,Z)} \right)^{p(E_0,Z)} \right]
\end{equation}
with $E_0$ the PE energy, $d$ the film thickness, $R$ the range and $p$ the transmission parameter. The transmission curves normalized this way can be characterized by the transmission parameter $p$. In figure \ref{fig:10c}, the transmission curves with different $p$-values are plotted as well using eq. \eqref{eq:15}. Lighter elements have a transmission characteristic similar to the curve with $p \approx 2$, while heavier elements have curves with $p \approx 1.5$. The $p$-value is constant over a wide range of energy $E_0$ for heavier elements, but depends on $E_0$ for lighter elements. For aluminum and alumina, it is fairly constant and $p \approx 1.9$ for $E_0=\SI{2}{keV}$ to $\SI{10}{keV}$. 

In figure \ref{fig:10c}, the transmission parameter is determined by superimposing curves calculated with eq. \eqref{eq:15} onto the normalized measurement data. The reduced film thickness $d/R$ is determined for each film thickness $d$ by using eq. \eqref{eq:5} to calculate the range $R$ for each $E_0$. The thicker film with $d=\SI{58.6}{\nm}$ has a transmission characteristic as predicted by eq. \eqref{eq:15} for alumina with $p \approx 1.8$, while the p-value and characteristics deviates for the thinner films. For the films with $d=\SI{13.8}{\nm}$ and $d=\SI{33.6}{\nm}$, the transmission parameters are $p\approx1.3$ and $p\approx1.5$, which resembles the transmission characteristics of PEs passing through gold and silver foils, respectively.

Hence, the transmission characteristic of ultra-thin films depends on the film thickness. One of our goal is to optimize the film thickness, so that the tynodes can perform optimally for sub-2 \SI{}{\keV} electrons. For this energy range, the transmission parameter $p(E_0,Z)$ is energy-dependent for alumina \cite{Fitting1974}. As we have shown, the transmission parameter is lower for the thinner films in comparison, which indicates that a relatively larger fraction of PEs are absorbed within the film. This is beneficial to (transmission) secondary electron emission since more energy is transferred and might be one of the contributing factors to the high TEY of the thinnest membrane.

%%%%%%%%%%%%%%%%%%%%%%%%

\subsection{Transmission secondary electron yield}
\label{ssec:normalized}

\begin{figure}[b!]
  \centering
	\subfloat[\label{fig:11a}] {\includegraphics[scale=0.135]{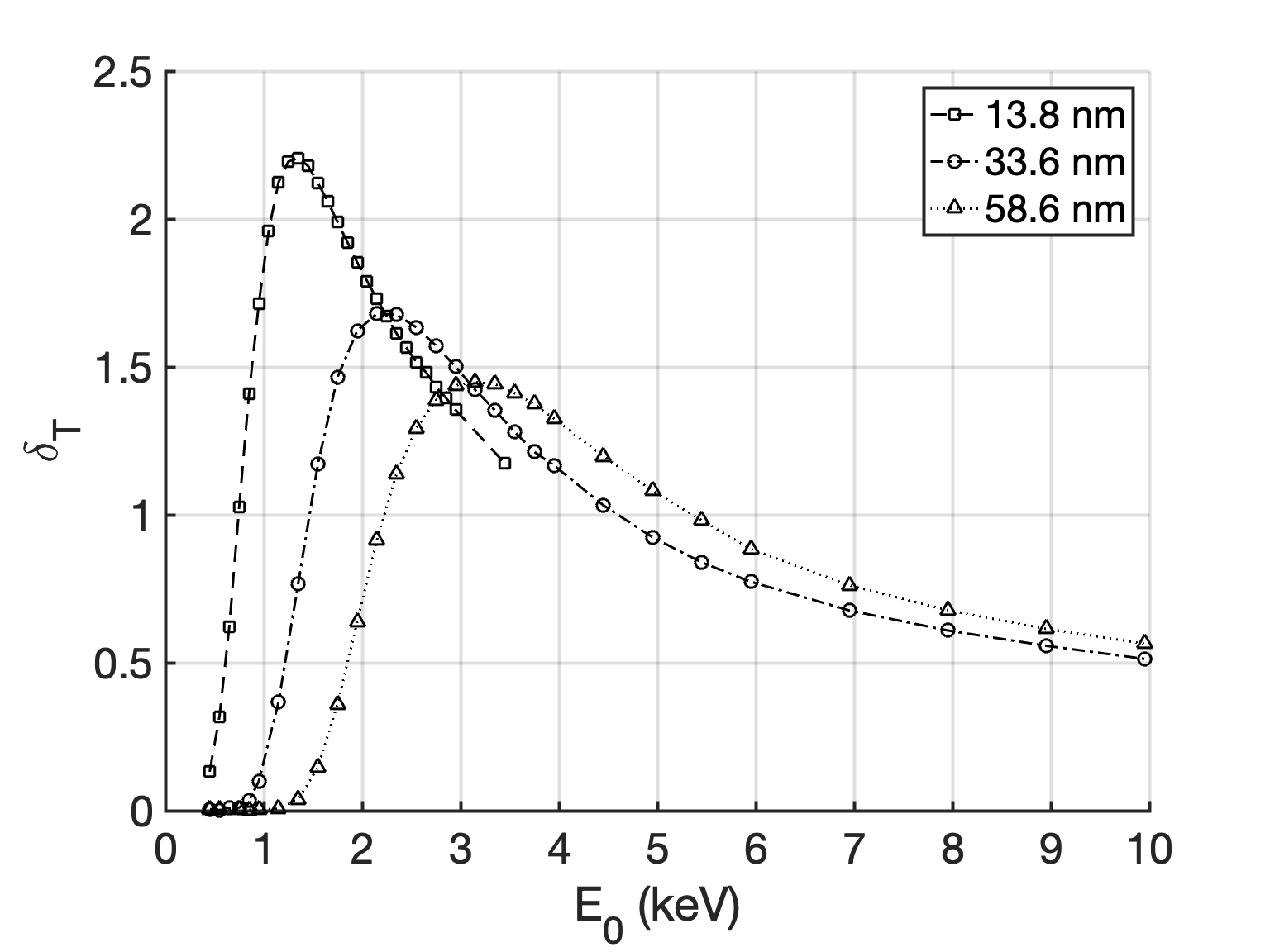}}
	\subfloat[\label{fig:11b}]{\includegraphics[scale=0.135]{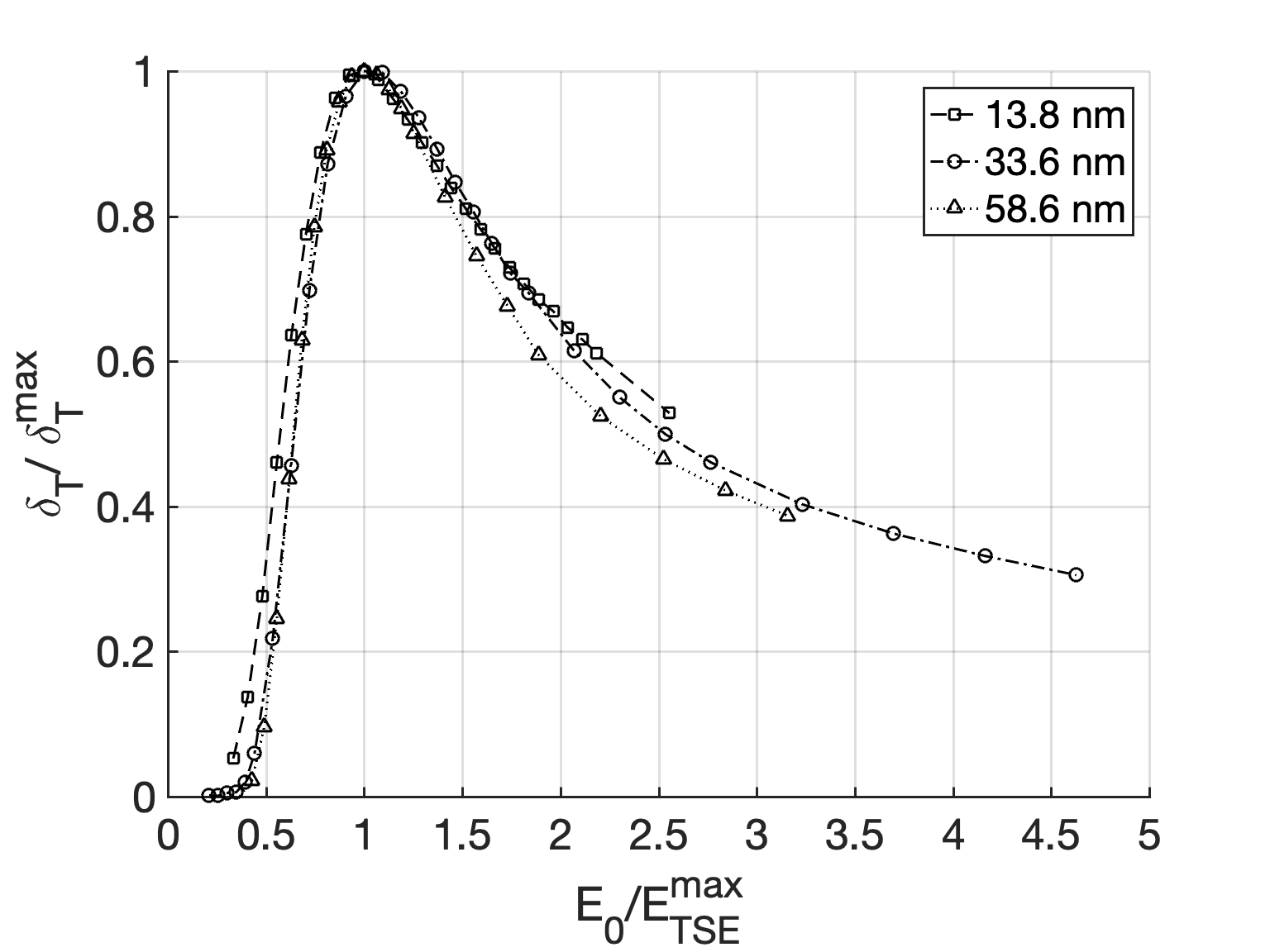}}\\
     \caption{\protect\subref{fig:10a}  TSEY \protect\subref{fig:10b} Normalized TSEY curve}
\end{figure}

The TSEY of thin films depends on its thickness. In figure \ref{fig:11a}, the TSEY curves for the bi-layer films are combined in one graph. Obviously, the film thickness determines the threshold energy $E_\text{th}$ at which the first TSEs are observed and the maximum energy $E_0^\text{max}$ at which the maximum TSEY is achieved. The width of the peak is narrower for the thinnest film and broadens for increasing thicknesses. Also, the maximum TSEY decreases as the film thickness increases. As discussed in \ref{ssec:versus}, the film with $d=\SI{13.8}{\nm}$ is near the optimal thickness. When the film thickness is further reduced, the interaction volume for the PEs will decrease and less SEs will be generated. 

The differences in the magnitude of the TSEY curve is solely due to the energy transfer process from PEs to the film. In the three-step model, the transport and escape mechanisms for internal SEs is the same for the films with different thicknesses. They consist of the same material and have the same surface condition. As such, the TSEY is proportional to the energy transfer near the exit surface of the films. The energy transfer profile $dE/dx(x,E_0)$ of an electron in bulk material can be determined by using this method; the energy transfer at depth $x$ is 'probed' by measuring the TSEY of a film with thickness $x$. By combining the results of multiple films with increasing thicknesses, the energy transfer profile in a solid can be obtained \cite{Fitting1977}. 

In our case the film thickness is fixed, while the electron energy $E_0$ increases. The TSEY curve in this case is 'probing' the increasing interaction volume of the PEs as the energy is increased. For simplicity, if we assume that the interaction volume is spherical or ellipsoidal and only grows in size for increasing energy $E_0$, then for $E_\text{th}$ the interaction volume will be a sphere (or ellipsoid) with a diameter equal to the film thickness. For $E_0^\text{max}$, the interaction volume will coincide with an interaction volume that is a half of a sphere with a diameter twice the film thickness. Using this simplified model, the width of the TSEY curves in figure \ref{fig:11a} coincides with the growth of the energy transfer profile as a function the electron energy $E_0$. The width of the TSEY curve is proportional to the film thickness. 

In figure \ref{fig:11b}, a normalization is applied to both the TSEY $\delta_T /\delta_T^\text{max}$ and the energy $E_0/E_0^\text{max}$. After normalization, the TSEY curves show remarkable resemblance. A similar result was found for carbon foils by Hölzl \& Jacobi \cite{Holzl1969}. There is a clear correlation between the threshold energy $E_\text{th}$ and the maximum energy $E_0^\text{max}$. By using the latter as the normalization constant, the normalized TSEY curve no longer depends on the film thickness. 

\begin{figure}
\centering 
\includegraphics[width=0.7\textwidth,origin=c,angle=0]{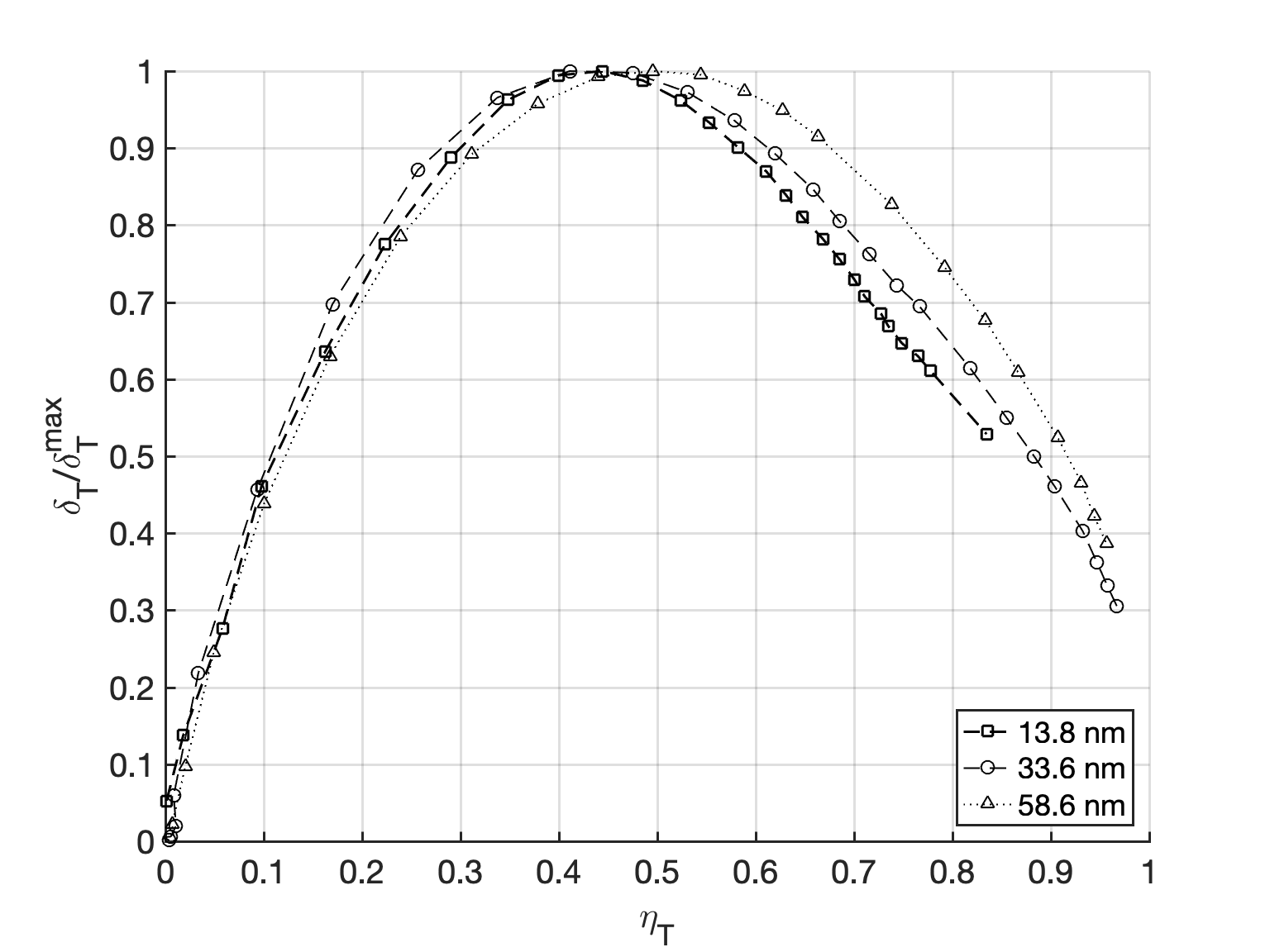}
\caption{\label{fig:12} Normalized yield vs. transmitted fraction}
\end{figure}

In figure \ref{fig:12}, the reduced TSEY $\delta/\delta_T^\text{max}$ is plotted against the transmitted fraction $\eta_T$, since they are strongly correlated. The max TSEY coincides with a transmitted fraction of approximately 0.4 to 0.5. As such, when PEs with the optimal energy $E_0^\text{max}$ is used to target the film, half of the initial PEs are either reflected or transmitted. Transmitted electrons still carry a considerable amount of energy: $E_x(x=R) \approx$ (0.3 to 0.4) $E_0$ \cite{Fitting2004}, which can induce tertiary currents and/or feedback signals in detectors such as the TiPC. They should therefore not be neglected in the detector design.

\section{Conclusions \& outlook}
\label{sec:con}

We have successfully constructed multilayered Al\textsubscript{2}O\textsubscript{3}/TiN membranes that can be used as tynodes in photodetectors. Two types of films have been made, a bilayer TiN/Al\textsubscript{2}O\textsubscript{3} and a tri-layer Al\textsubscript{2}O\textsubscript{3}/TiN/Al\textsubscript{2}O\textsubscript{3}. The tri-layer film has the conductive TiN encapsulated in order to improve the reliability of the manufacturing process. The TiN layer provides in-plane conductivity to sustain prolonged electron emission and to prevent charge-up. The highest TEY was achieved for the thinnest tri-layer film, which has a TEY of 3.1 (1.55 keV) for a film with a thickness of \SI{13.8}{\nm}. The requirements for the tynodes in TiPC is a TEY of 4 or higher for sub-2 \SI{}{\keV} electrons .

The results so far are promising and the tri-layer membrane design provides a solid foundation for future work. First, the tri-layer membrane samples presented in this work will be used to build a multi-stack prototype TiPC. In a recent publication, the electron emission from one tynode similar to the ones presented in this paper was measured in a dedicated vacuum setup using a TimePix chip as readout \cite{VanDerReep2020}. 

Second, the transmission yield of the tri-layer membrane can be improved by applying surface treatment, such as caesiation or hydrogen-termination. For this purpose, film-on-bulk samples can be used to measure the effect of termination on the reflection yield. However, a dedicated ultra-high vacuum system is required to prevent surface contamination. 

Third, new materials can be considered as substitute for ALD Al\textsubscript{2}O\textsubscript{3}. The fabrication process presented in this work can be easily adapted for new materials, such as ALD MgO, which is a promising candidate. The first experiments on film-on-bulk samples have shown a higher reflection yield compared to Al\textsubscript{2}O\textsubscript{3} \cite{Prodanovic2018b}. 

Lastly, the active surface area of tynodes can be increased by forming a corrugated membrane of ALD Al\textsubscript{2}O\textsubscript{3}/TiN/Al\textsubscript{2}O\textsubscript{3} film. The corrugated film has enhanced mechanical strength, which allows it to span over a larger surface area. This will improve the collection efficiency of the Timed Photon Counter.

\acknowledgments
We would like to thank H. Akthar, C. Hansson, S. Tao and J. Smedley for their contribution to the MEMBrane project. We are also thankful for the technical support we received from O. v. Petten (Nikhef), J. v.d. Cingel, H. v.d. Linden and C.Th. Heerkens on the experimental setup. We are grateful to the Else Kooi Lab that provided the training and facilities to manufacture the films. This work is supported by the ERC-Advanced Grant 2012 MEMBrane 320764.

\bibliographystyle{IEEEtran}
\bibliography{document.bbl}

%%%%%%%%%%%%%%%%%% Appendix %%%%%%%%%%%%%%%%%%

\appendix
\appendixpage
\addappheadtotoc

\section{Sample geometry correction}
\label{sec:samplecor}

The reduction in yield due to reabsorption by the walls of the window frame is estimated by comparing two measurements on a p-type silicon membrane with a thickness of 39 nm. The width, height and aspect ratio of the window is the same as the other samples presented in this paper. The emission surface of the silicon membrane is identical on both side, so the reduction in yield is only due the window frame, which obstruct electron emission on one side. 

\begin{figure}
  \centering
	\subfloat[\label{fig:13a}] {\includegraphics[scale=0.2]{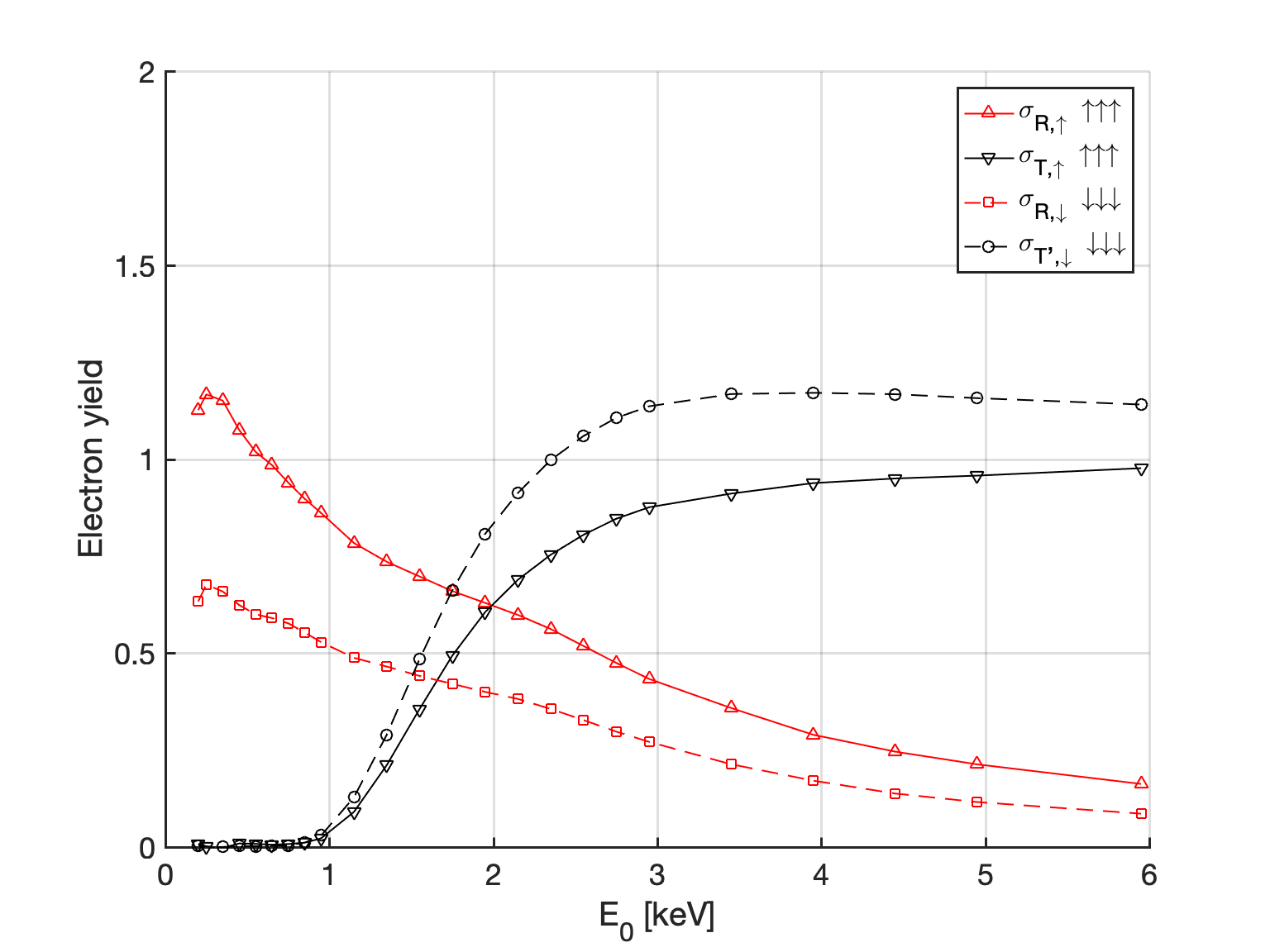}}\\
	\subfloat[\label{fig:13b}]{\includegraphics[scale=0.2]{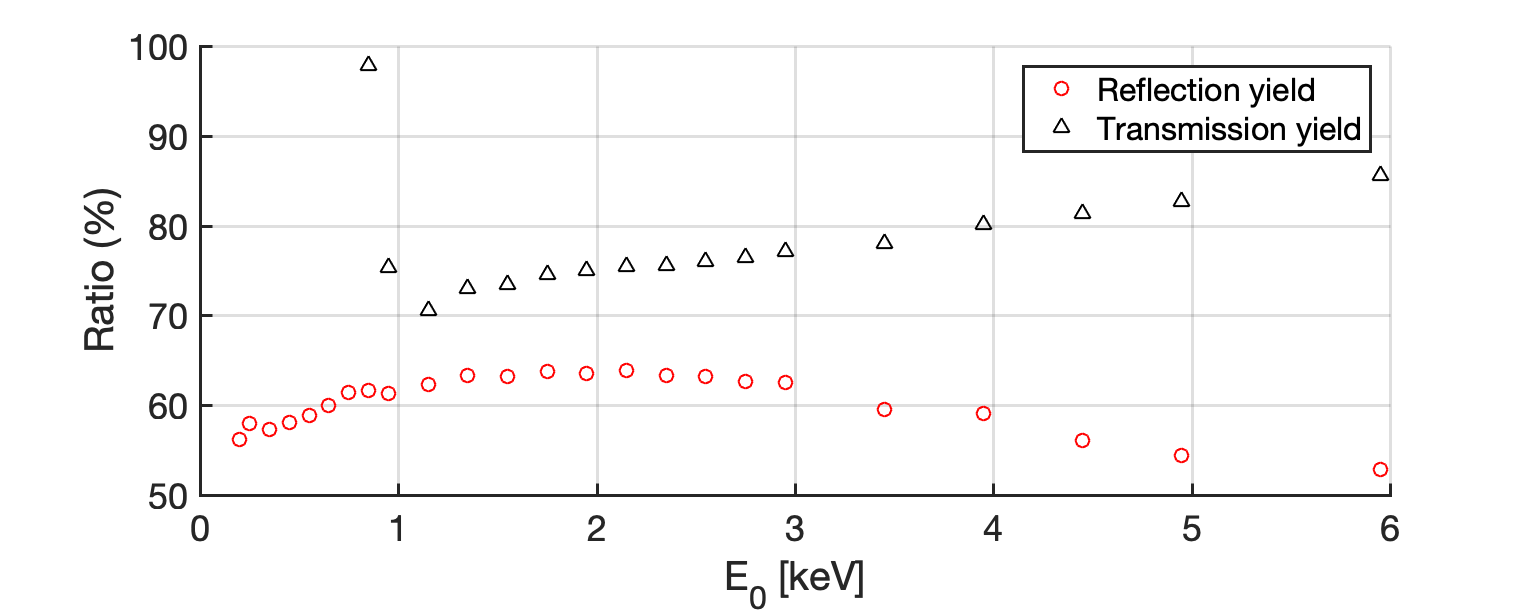}}
     \caption{\protect\subref{fig:13a} The influence of reabsorption by the window opening on secondary electron emission of a silicon membrane on a SOI substrate. The arrows indicate the direction the flat side is facing. \protect\subref{fig:13b} The ratio between the obstructed and unobstructed yield for REY and TEY.}
\end{figure}

In figure \ref{fig:13a}, the REY and TEY curves of two separate measurements are given. The first measurement, with the window frame facing upwards, shows a reduction in REY, while the TEY is unaffected. In the second measurement, the window frame is facing downwards and the result shows a reduction in the TEY, while the REY is unaffected. The ratio between the reduced and unaffected yields is given in figure \ref{fig:13b}. Reabsorption decreases the REY by 35 to 45\% and the TEY by 15 to 30\%. A correction can be applied to the obtained results, but the correction factor depends on the beam energy and is only valid for these specific samples.

\section{Measurement setup correction}
\label{sec:meascor}

When the measurement setup is operated with a positive sample bias, a forward scattered electron (FSE) can cause tertiary currents on the retarding grid and collector. As a result, the transmitted fraction is lower than expected: it converges to 0.8, while it should converge to 1 for high energetic PEs. A correction term is therefore needed to account for the tertiary currents caused by the FSEs on the retarding grid and collector. In this appendix, we estimate the correction term by considering the scenarios in which a FSE can induce tertiary currents. For simplicity we will use the following assumptions in the model:
 
 \begin{itemize}
  \item Each FSE scatters only once. 
  \item The electron emission distributions of the SEs and BSEs are uniform, hemispherical and normal to the surface.
\end{itemize}

In the first scenario, a FSE scatters on the surface of the retarding grid. It can either be scattered towards the collector, be absorbed by the grid or be backscattered towards the sample holder. In the last case, the BSE current $I_{\text{BSE,grid}}$ can be considered as a 'tertiary' current, since the FSE will be recollected by the sample holder and not be counted. In all cases, tertiary electrons (unwanted SEs) are generated that will either flow to the collector or the sample holder since both are positively biased with respect to the grid. This induces a tertiary current $I_{\text{tertiary}}$ from the grid to the sample holder. The tertiary current from the grid to the collector can be disregarded, since it has a zero net effect on the transmitted fraction:  $\eta_T(E_0)=(I_{RG+}-I_{\text{tertiary,c}})+(I_{C+}+I_{\text{tertiary,c}}$). 

In the second scenario, a FSE scatters on the wall of the collector and will either be absorbed or backscattered. Again, the BSE current $I_{\text{BSE,col}}$ is considered as a 'tertiary' current. In both cases, tertiary electrons are generated as well, but they will not be able to escape due to the positive bias of the collector and does not cause a tertiary current.   
	
If we include the tertiary currents, then the measured transmitted fraction is given by:
\begin{subequations}\label{eq:16}
\begin{align}
\label{eq:16:1}
\eta_{T,\text{meas}}(E_0) &=  \frac{I_{T}-I_\text{BSE,grid}-I_\text{BSE,col}-I_{\text{tertiary}}}{I_0}
\\
\label{eq:16:2}
 &=  \frac{I_T}{I_0}-\frac{I_\text{BSE,grid}+I_\text{BSE,col}+I_{\text{tertiary}}}{I_0}
\\
\label{eq:16:3}
 & = \eta_{T,\text{true}}(E_0)) - \frac{I_\text{BSE,grid}+I_\text{BSE,col}+I_{\text{tertiary}}}{I_0}
\\
\label{eq:16:4}
\eta_{T,\text{meas}}(E_0) &= \eta_{T,\text{true}}(E_0)- \alpha(E_0)
\end{align}
\end{subequations}
with $I_T$ the (true) transmission current in \SI{}{\nano\ampere}, $\eta_{T,\text{true}}(E_0)$ the true transmitted fraction, $I_\text{BSE,col}$ the BSE current from the collector to the sample holder in \SI{}{\nano\ampere}, $I_\text{BSE,grid}$ the BSE current from the grid to the sample holder in \SI{}{\nano\ampere}, $I_\text{tertiary}$ the tertiary current from the grid to the sample holder in \SI{}{\nano\ampere} and $\alpha(E_0)$ a correction term. %

The tertiary current from the grid is given by:
\begin{equation}
\label{eq:17}
I_\text{tertiary} = I_0\gamma\delta_{R,\text{grid}}
\end{equation}
with $I_0$ the primary current, $\gamma$ the opacity of the retarding grid mesh and $\delta_{R,\text{grid}}$ the reflection yield of the grid mesh material. 
The BSE current from the grid is given by:
\begin{equation}
\label{eq:18}
I_\text{BSE,grid}  = I_0\gamma\varepsilon_\theta\eta_{R,\text{grid}}
\end{equation}
where $I_0$ is the primary current, $\gamma$ the opacity of the retarding grid mesh, $\varepsilon_\theta$ the backscattered angle efficiency and $\eta_{R,\text{grid}}$ the BSE yield of the retarding grid material. 
The BSE current from the collector is given by:
\begin{equation}
\label{eq:18}
I_\text{BSE,col}  = I_0(1-\gamma)^2\varepsilon_\theta\eta_{R,\text{col}}
\end{equation}
where $I_0$ is the primary current, $\gamma$ the opacity of the retarding grid mesh, $\varepsilon_\theta$ the backscattered angle efficiency and $\eta_{R,\text{col}}$ the BSE yield of the collector material. In this case, a BSE only counts towards the current if it passes through the retarding grid again. Otherwise, we assume that it remains trapped between the collector and grid, and will eventually be absorbed. Therefore, the probability for a BSE to pass the grid twice is given by the square of the transparency of the grid: $(1-\gamma)^2$.

The backscattered angle efficiency $\varepsilon_\theta$ is the ratio of the field of view of an electron and its emission angle distribution. The field of view of an electron on the collector wall is dome-shaped. We assumed that SEs and BSEs are emitted uniformly and hemispherically. Therefore, we can use the ratio of the surface area of a spherical cap (of a dome) to a hemisphere to find $\varepsilon_\theta$. The surface area of a spherical cap is given by $A_\text{cap}= 2 \pi r^2\times(1-\cos{\theta})$. The surface of this hemisphere (without its base) is given by $A_\text{hemi}=2\pi r^2$. The polar angle of an electron at the bottom center of the collector is $\theta=\frac{\pi}{4}$. As a result, the backscattered angle efficiency is given by $\varepsilon_\theta=(1-\cos{\theta}=0.29)$. 

The correction term $\alpha(E_0)$ is given by:
\begin{subequations}\label{eq:19}
\begin{align}
\label{eq:19:1}
\alpha(E_0) &=   \frac{I_\text{BSE,grid}+I_\text{BSE,col}+I_{\text{tertiary}}}{I_0}= \frac{I_0\gamma\delta_{R,\text{grid}}+I_0\gamma\varepsilon_\theta\eta_{R,\text{grid}}+I_0(1-\gamma)^2 \varepsilon_\theta \eta_{R,\text{col}}}{I_0}
\\
\label{eq:19:2}
{}& = \gamma\delta_{R,\text{grid}}+\gamma\varepsilon_\theta\eta_{R,\text{grid}}+(1-\gamma)^2\varepsilon_\theta\eta_{R,\text{col}}
\\
\label{eq:19:3}
{}&= 0.1\times1.3+0.1\times{0.29}\times0.33+(1-0.1)^2\times{0.29}\times0.34\approx0.22
\end{align}
\end{subequations}
The correction term is estimated for PEs with $E_0=\SI{10}{\keV}$. The retarding grid has an opacity of 0.1 and is made of stainless steel. The RSE and BSE yield for $E_0=\SI{10}{\keV}$ are $\delta_{R,\text{grid}} =1.3$ and $\eta_{R,\text{grid}}=0.33$, respectively. The BSE yield of the copper collector is $\eta_{R,\text{col}}=0.34$. 

Despite the simplicity of the model, the estimated correction term $\alpha(\SI{10}{keV})\approx0.22$ is close to the expected value. However, the correction term is energy-dependent. For low electron beam energy, the PEs will lose a large fraction of its energy before it can transmit through the membrane. As such, they will barely induce tertiary currents and the correction term will be close to zero. On the other end, high energetic PEs will barely lose energy as they transmit through the membrane. It is likely that they will scatter multiple times within the closed collector system and induce tertiary currents on each impact. Therefore, we can assume that tertiary currents are directly proportional to the electron beam energy. As a result, we can make a rough estimation for the correction term $\alpha(E_0)$: when the electron beam energy is increased from $E_c$ to $E_0=\SI{10}{\keV}$, the correction term increases linearly from 0 to 0.22.

\end{document}